\documentclass{article}
\usepackage{graphicx}  
\usepackage{amsmath}   
\usepackage[compress]{cite}
\usepackage{amssymb}   
\usepackage{bm} 
\usepackage{dcolumn}
\usepackage{color}
\usepackage{mathrsfs}
\usepackage[utf8]{inputenc}
\usepackage{amsfonts}
\usepackage{varioref}
\usepackage{color}
\usepackage{subfig}
\usepackage{float}
\usepackage{booktabs,multirow,array}
\usepackage[table]{xcolor}
\usepackage{float}
\RequirePackage[colorlinks,citecolor=blue,urlcolor=magenta,linkcolor=blue]{hyperref}
\def\EH{Einstein-Hilbert }
\def\LL{Lanczos-Lovelock }
\def\EGB{Einstein-Gauss-Bonnet }
\def\scc{strong cosmic censorship conjecture}

\def\gr{general relativity}
\def\RN{Reissner-Nordstr\"{o}m }

\def\ch{Cauchy horizon}

\def\coh{cosmological horizon }
\def\qnm{quasi-normal }
\def\PS{photon sphere }
\def\NE{near extremal }
\def\DS{de Sitter }

\labelformat{section}{Section #1} 
\labelformat{subsection}{Section #1} 
\labelformat{subsubsection}{Section #1}
\labelformat{subsubsubsection}{Section #1}
\labelformat{equation}{Eq.~(#1)} \labelformat{figure}{Fig.~#1} 
\labelformat{subfigure}{Fig.~\thefigure#1} 
\labelformat{table}{Table.~#1} 
\labelformat{appendix}{Appendix #1}
\title{Strong cosmic censorship conjecture in higher curvature gravity}
\author{Akash K Mishra\footnote{akash.mishra@iitgn.ac.in}$~^{1}$ and
Sumanta Chakraborty\footnote{sumantac.physics@gmail.com}$~^{2,3}$ \\
{\small{$^{1}$ Indian Institute of Technology, Gandhinagar-382355, Gujarat, India}}\\
{\small{$^{2}$ School of Physical Sciences, Indian Association for the Cultivation of Science, Kolkata-700032, India}}\\
{\small{$^{3}$ School of Mathematical and Computational Sciences}}\\
{\small{Indian Association for the Cultivation of Science, Kolkata-700032, India}}}
\date{ }  
\begin{document}
 
\maketitle
\begin{abstract}
Deterministic nature of general relativity is ensured by the strong cosmic censorship conjecture, which asserts that spacetime cannot be extended beyond Cauchy horizon with square integrable connection. Although this conjecture holds true for asymptotically flat black hole spacetimes in general relativity, a potential violation of this conjecture occurs in charged asymptotically de Sitter spacetimes. Since it is expected that Einstein-Hilbert action will involve higher curvature corrections, in this article we have studied whether one can restore faith in the strong cosmic censorship when higher curvature corrections to general relativity are considered. Contrary to our expectations, we have explicitly demonstrated that not only a violation to the conjecture occurs near extremality, but the violation appears to become stronger as the strength of the higher curvature term increases.  
\end{abstract}
\section{Introduction and Motivation}

The Strong Cosmic Censorship Conjecture, broadly speaking, asserts that all the physically reasonable solutions of the Einstein's equations with regular initial data are globally hyperbolic, which in turn implies that \gr\ is deterministic in nature \cite{hawking_ellis_1973,Wald:1984rg}. However, the existence of a \ch\ in several realistic solutions of Einstein's equations, may indicate a possible violation of \scc, since a \ch\ is regarded as the boundary of maximum Cauchy development of an initial data given on a Cauchy hypersurface. Therefore the breakdown of \scc\ or equivalently understanding the deterministic nature of the theory boils down to the question, whether the spacetime can be extended beyond the \ch. If the metric is regular at the \ch, it is possible to construct a geodesic that can be extended beyond the \ch\ into regions where any further evolution of the geodesic cannot be uniquely obtained from the initial data \cite{Costa:2014yha}. This scenario can be considered as a potential violation of the strong cosmic censorship conjecture. One possible resolution to this problem, as proposed by Penrose, has to do with the unstable nature of the \ch\ with respect to any small perturbation \cite{Simpson1973}. More precisely, if the perturbations at the \ch\ grow unboundedly, then ultimately they will turn into a curvature singularity and hence the problem of crossing the \ch\ can be avoided. This process of turning a \ch\ to a curvature singularity is known as the mass inflation in the literature \cite{PhysRevD.41.1796,PhysRevLett.67.789,poisson_2004,Bhattacharjee:2016zof}. This can also be understood from the fact that all the incoming waves will be blue shifted by an infinite amount as they approach the \ch, leading to existence of a singularity. This feature will survive for asymptotically flat spacetimes, where the perturbations have a power-law decay at late times and hence the exponential growth $\Phi \sim e^{\kappa_{-} u}$, with $\kappa_{-}$ being the surface gravity at the Cauchy horizon, always dominates, leading to singular behaviour. 

On the other hand, for asymptotically de Sitter spacetimes, the perturbations at late times also decay exponentially, which has the possibility of being cancelled by the exponential growth at the \ch, leading to extension of the spacetime beyond the \ch\ \cite{PhysRevD.5.2419,Angelopoulos:2016wcv}. More precisely, for the case of asymptotically de Sitter spacetimes, e.g., Reissner-Norsdr\"{o}m-de Sitter black hole, the perturbation attains an exponentially decaying late-time tail $\Phi \sim e^{-\alpha u}$, where $\alpha = -\text{Im}(\omega)$ is the spectral gap related to the lowest-lying \qnm frequency. Therefore it is indeed possible, at least for a certain range of parameters, where the exponential decay of perturbation is balanced by the exponential growth of perturbation at the \ch, thus avoiding any mass inflation singularity. Hence the quantity of interest in such an analysis is the relative strength between the decay and growth of the perturbation at the \ch, which is determined by the quantity $\beta \equiv (\alpha/\kappa_{-})$. It so happens that for $\beta>1/2$, the late-time decay of the perturbation becomes strong enough to overcome the growth at \ch, thereby leading to a violation of \scc\ \cite{PhysRevLett.120.031103}. In the absence of a general proof for \scc, such an analysis plays a very crucial role in order to test the conjecture, i.e., by looking for possible counterexamples. This approach have been used recently by several authors in order to test the validity of \scc\ conjecture for \gr\ on various asymptotically de Sitter black hole spacetimes in four and higher dimensions with different test fields \cite{Mo:2018nnu,Dias:2018ufh,Ge:2018vjq,Rahman:2018oso,Destounis:2018qnb,Gwak:2018rba,Gim:2019rkl,Liu:2019lon,Rahman:2019uwf,Guo:2019tjy,Dias:2019ery,Liu:2019rbq,Cardoso:2018nvb,Dias:2018ynt,Dias:2018etb}. The central result arising out of these analyses is the realization that the \scc\ is violated in the near extremal regime for non-rotating black holes, while for rotating black holes, the violation can be avoided. For black holes in Born-Infeld-\DS and Horndeski theory, the \scc\ has also been recently studied in \cite{Gan:2019jac,Destounis:2019omd}.

Surprisingly, no such analysis for validity of \scc\ has ever been extended to black holes in higher curvature gravity theories. Even though general relativity describes the gravitational interaction around us very nicely, it also has several shortcomings. The most notable among various ones are the singularity problem and late time acceleration of the universe. Besides there are also numerous other motivations for looking for gravity theories beyond \gr, including non-renormalizability of the gravitational action \cite{doi:10.1063/1.1724264,PhysRevD.16.953,PhysRevD.36.392,Buchbinder:1992rb}. Thus it is reasonable to believe that \gr\ is only an effective field theory, which must be supplemented by higher curvature corrections at strong gravity regime. The most natural generalization of the Einstein-Hilbert action involving higher curvature corrections is the \LL gravity, containing at most second derivative of the metric \cite{Lovelock:1971yv,ZWIEBACH1985315,PhysRevLett.55.2656,ZUMINO1986109,PADMANABHAN2013115,Chakraborty:2015wma}. This motivates us to study whether the \scc\ holds in the presence of higher curvature terms. There is a hope that even though the \scc\ is violated for certain solutions in \gr, when the higher curvature corrections are taken into account it may be respected. 

Following which, in this work, we have studied the \scc\ in the context of two higher curvature theories. As our first example, we consider the case of \EGB gravity in five and higher spacetime dimensions, which is the second-order term of the \LL Lagrangian. The \EGB theory admits spherically symmetric charged black hole solutions with a cosmological constant \cite{Cai:2003gr,Guo:2018exx,Cai:2013qga,EslamPanah:2017yoc,Wei:2012ui,Chakraborty:2015taq}, which involves a \ch. Thus one can ask whether the solution can be extended beyond the \ch. Our second example involves the study of pure lovelock black hole solutions \cite{PhysRevD.100.084011,Dadhich:2015nua,Dadhich:2016fku,Chakraborty:2016qbw,Dadhich:2015ivt,Gannouji:2013eka} in dimensions $d\geq (3k+1)$, with `k' being the lovelock order, i.e., $k=1$ is the pure Einstein Gravity, while $k=2$ is pure Gauss-Bonnet Gravity and so on. We would like to emphasize that, although the pure lovelock solutions may not represent a physical black hole, it does provide a natural platform to study the effect of higher curvature terms to the \scc, which is the ultimate aim of our work. 

The article is arranged as follows: In \ref{Section 2} we start by reviewing the relationship between the \qnm frequency of the \PS modes and the Lyapunov exponent associated with the photon sphere. This is a general result for any spherically symmetric spacetime irrespective of the underlying theory of gravity. Subsequently in \ref{Section 3} we present a detailed analysis for obtaining the \qnm frequency for \EGB black holes numerically and demonstrate the violation of \scc. The above procedure has been repeated in order to obtain the \qnm frequencies for pure Lovelock black holes in an appropriate spacetime dimensions and show the violation of \scc\ in \ref{Section 4}. We have compared this violation with the corresponding scenario in Einstein gravity to illustrate the effect of higher curvature terms. From both of our examples, we conclude that the violation of \scc\ becomes even stronger when higher curvature terms are added. We end with a brief discussion and possible future outlooks in \ref{Section 5}.

\emph{Notations and Conventions:} We have set the fundamental constants $c=1=G$. The Roman indices $(a,b,c,\cdots)$ are used to denote spacetime indices. The Greek indices $(\mu,\nu,\alpha,\cdots)$, on the other hand, are used to denote spatial indices on a spacelike hypersurface.
\section{Strong cosmic censorship conjecture and Quasi-normal modes: A brief overview}\label{Section 2}

The stability of black holes under small perturbation, one of the most important area of research in black hole physics, requires the computation of the \qnm modes. These modes are the eigenfunctions of the perturbation equation with respect to some special set of boundary conditions, i.e., only ingoing modes at the event horizon and outgoing modes at infinity. The real part of the associated eigenvalues, known as \qnm mode frequencies, determines the time period of oscillation, while the imaginary part dictates the decay rate of the perturbation. It is the decay rate of the perturbation, which is central to the stability of a black hole. Thus the question of stability of a black hole spacetime is linked with the sign of the imaginary part of the \qnm mode frequency, $\omega$. For most black hole spacetimes, because of the complex structure of the perturbation equation, it is a daunting task to obtain an analytical expression for the \qnm frequencies by solving the perturbation equation. A relatively more straightforward task is to obtain the \qnm mode frequencies by solving the perturbation equation numerically, and various numerical techniques have been developed over the last few decades to compute the \qnm mode frequencies accurately. However, in certain limiting cases, it is indeed possible to obtain an analytical expression of the \qnm mode frequency. One such limiting case is the ray optics approximation or the eikonal limit, where both the real and imaginary part of the \qnm mode frequency is related to various geometric constructs associated with the photon sphere. This stems from the fact that the effective potential experienced by a photon in a black hole spacetime is identical to the potential experienced by a test field in this black hole spacetime, in the large angular momentum limit. Thus the \qnm mode frequencies, which are directly connected with the potential in the perturbation equation in the eikonal limit (this also corresponds to large angular momentum limit), gets related to the potential a photon experiences. In particular, the imaginary part of the \qnm mode frequency is related to the Lyapunov exponent associated with the instability of the photon sphere and the real part to the angular velocity of the photon sphere \cite{PhysRevD.79.064016,PhysRevD.80.064004,PhysRevD.31.290,Cornish:2003ig,Bombelli_1992,PhysRevLett.52.1361}, such that
\begin{equation}\label{imomega}
\omega_{n}=\Omega_{\rm ph}\ell-i\left(n+\frac{1}{2}\right)\lambda_{\rm ph}~,
\end{equation}
where $\ell$ is the angular momentum and $n=0,1,2....$ represents the overtone number. We also refer the reader to \cite{Konoplya:2017wot,Konoplya:2019hml}, where the above correspondence is discussed in the context of higher curvature gravity.

The Lyapunov exponent $\lambda_{\rm ph}$, associated with the photon sphere, determines the rate at which a geodesic located at the photon sphere diverge or converge with respect to a nearby geodesic. Further, $\Omega_{\rm ph}$ is the angular velocity of a photon located at the maxima of the photon sphere. In a d-dimensional static and spherically symmetric spacetime one can explicitly write down the expressions for $\lambda_{\rm ph}$ and $\Omega_{\rm ph}$ in terms of the metric coefficients, which reads 
\begin{equation}\label{generalsphsymm}
ds^{2}=-f(r)dt^{2}+f(r)^{-1}dr^{2}+r^{2}d\Omega^{2}_{d-2}~,
\end{equation} 
Exploiting the fact that the spacetime posses spherical symmetry, it is convenient to restrict our attention only to the equatorial plane, which is identified by setting the azimuthal angels to $\pi/2$ and hence the Lagrangian associated with the geodesic motion takes the form,
\begin{equation}\label{lagrangian}
L=\frac{1}{2}\left(-f(r)\dot{t}^{2}+f(r)^{-1}\dot{r}^{2}+r^{2}\dot{\phi}^{2}\right)
\end{equation}
where `dot' represents derivative with respect to the affine parameter. Here $t$ and $\phi$ are cyclic coordinates and correspondingly the energy and angular momentum, $p_t =-E$ and $p_\phi = L$ are the constants of motion. The unstable circular null trajectory or, more commonly, the photon sphere, is determined by the equations $V'_{\rm eff}(r_{\rm ph})=V''_{\rm eff}(r_{\rm ph})=0$, where $V_{\rm eff}$ is the effective potential a photon experiences. These equations further reduce to,
\begin{equation}\label{PS_eqn}
\begin{aligned}
\frac{E^{2}}{L^{2}}&=\frac{f(r)}{r^{2}}
\\
2f(r)&=rf'(r)
\end{aligned}
\end{equation}
The Lyapunov exponent is determined by taking the variation of the effective potential $V_{\rm eff}$ as $r\rightarrow r_{\rm ph} + \delta r$ and hence one can show the following time evolution, $\delta r\sim \exp(\pm \lambda_{\rm ph} t)$, where the Lyapunov exponent $\lambda_{\rm ph}$ has the following expression in terms of the effective potential \cite{PhysRevD.79.064016},
\begin{equation}\label{lyapunov}
\lambda_{\rm ph}=\sqrt{\dfrac{V_{\rm eff}''}{2\dot{t}^{2}}}\bigg|_{r=r_{\rm ph}} = \sqrt{\frac{f(r_{\rm ph})}{2}\left(\frac{2f(r_{\rm ph})}{r_{\rm ph}^{2}}-f''(r_{\rm ph})\right)}
\end{equation}
As per our convention the time dependence of the perturbation goes as $\exp({-i\omega_{n}t})$ and hence the imaginary part of the \qnm mode frequency must be negative ensuring stability. Since the longest lived \qnm mode frequency correspond to the $n=0$ mode in \ref{imomega}, the quantity of interest for strong cosmic censorship conjecture, i.e., $\beta\equiv \{-\textrm{min}~(\textrm{Im}~\omega_{n})/\kappa_{\rm ch}\}$, is given by,
\begin{equation}\label{generalbeta}
\beta_{\rm ph}=\frac{\lambda_{\rm ph}}{2\kappa_{\rm ch}}~=\frac{1}{2\kappa_{\rm ch}}\left\{\sqrt{\frac{f(r_{\rm ph})}{2}\left(\frac{2f(r_{\rm ph})}{r_{\rm ph}^{2}}-f''(r_{\rm ph})\right)} \right\}
\end{equation}
This finishes one part of the story, since \ref{imomega} yielding an analytical expression for \qnm mode frequencies holds true only in the large $\ell$ limit. Since the \qnm mode frequencies in this context solely depends on the photon sphere these are generally referred to as the \PS modes. However, in presence of the electromagnetic charge and cosmological constant, the \qnm spectrum of a black hole spacetime possesses two other characteristic \qnm modes, namely, the \DS modes and the \NE modes. 
{The \DS mode becomes relevant when the cosmological horizon lies far away from the event horizon, i.e., in the limit when the cosmological constant goes to zero. On the other hand, the \NE modes dominate the spectrum in the extremal limit, i.e., when the \ch\ approaches the event horizon. The \qnm mode frequency associated with a \DS mode can be solely determined from the asymptotic structure of spacetime, which has the following form \cite{PhysRevD.70.064024,LopezOrtega:2012vi,PhysRevD.66.104018,Abdalla:2005hu},
\begin{equation}\label{qnmds}
\omega _{n,\textrm{dS}}=-i\left(\ell+2n\right) \kappa _{\rm c}
\end{equation} 
where $\kappa_{\rm c}$ is the surface gravity associated with the cosmological horizon and the \qnm mode frequencies are purely imaginary, as evident from \ref{qnmds}. The minimum value for the imaginary part of the \qnm mode frequency corresponds to $\omega_{n=0,\textrm{dS}}=-i\ell \kappa_{\rm c}$. Note that, the analytical expressions presented in this section for the \PS and \DS modes are not exact and has been obtain with some approximation. Hence in the subsequent sections we compute the \qnm frequencies numerically which can be further compared with their corresponding analytical expressions given in this section. The final set of modes, which are of importance in this context are the near extremal modes. These appear when the \ch\ and the event horizon approach each other. In the context of black holes in \gr\ it was possible to provide an analytical estimation for these modes, however, in the present context it turns out to be difficult to write down analytical form for the near extremal modes due to complicated nature of the equation determining the event horizon. Hence we will not attempt to write down any analytical expression for the \NE modes, rather we will compute it numerically.

Violation of the \scc\ requires the existence of weak solutions across the Cauchy horizon. These weak solutions are said to exist, when the integral of the non-linear field equations multiplied with a smooth function over a region around the Cauchy horizon is finite. For the case of $m$th order Lovelock theory, the gravitational field equations will involve $m$ curvatures terms, and hence the existence of weak solutions will demand, 
\begin{equation}\label{qnmds}
\int _{\mathcal{V}}d^{d}x\sqrt{-g}~\psi \left(\partial \Gamma +\Gamma ^{2}\right)^{m}=\int _{\mathcal{V}}d^{d}x\sqrt{-g}~\psi \left(\partial \phi\right)^{2}~,
\end{equation} 
where, $\mathcal{V}$ is a volume around the Cauchy horizon and $\psi$ is a smooth function. Thus for the above integrals to exist, $\phi$ should be square integrable and hence must belong to $H^{1}_{\rm loc}$, while $\Gamma$ should be a function in $L^{2m}_{\rm loc}$ and thus $g_{\mu \nu}$ should be $H^{m}_{\rm loc}$. Thus for gravitational perturbation to extend weakly across the Cauchy horizon one requires stronger condition, than that for the scalar field. Since, extension of any perturbation, e.g., scalar, electromagnetic or gravitational, across the Cauchy horizon would constitute a violation of the \scc, it follows that if $\beta>(1/2)$ for scalar perturbation, then \scc\ will be violated. However, for gravitational perturbation, one must be careful, before analyzing the limit on $\beta$, which we leave for the future.

To see how the choice for $\beta$ can be related to the validity of \scc, consider the field equation $\square \Phi=0$ that a test scalar field obeys in the $d$ dimensional static and spherically symmetric spacetime. Due to existence of timelike and angular Killing vectors in the spacetime, the scalar field can be expressed as, $\Phi(t,r,\Omega)=e^{-i\omega t}R(r)h(\Omega)$, where $h(\Omega)$ corresponds to spherical harmonics associated with $(d-2)$ dimensional unit sphere. The function $R(r)$ satisfies a second order differential equation, whose two independent solutions, regular at the Cauchy horizon, reads
\begin{align}
\Phi^{(1)}(t,r,\Omega)=e^{-i\omega u}R^{(1)}(r)h(\Omega);\qquad \Phi^{(2)}(r)=e^{-i\omega u}R^{(2)}(r)(r-r_{\rm ch})^{i\omega_{n}/\kappa_{\rm ch}}h(\Omega)
\end{align}
Here $\omega_{n}$ is the \qnm mode frequency and $\kappa_{\rm ch}$ is the surface gravity associated with the Cauchy horizon. As a consequence, the integral of the kinetic term of the scalar field correspond to the integral of $(r-r_{\rm ch})^{2(i\omega_{n}/\kappa_{\rm ch}-1)}$, which in turn corresponds to, $(r-r_{\rm ch})^{2(\beta-1)}$. Here $\beta$ has already been defined above in terms of $\textrm{Im}~\omega_{n}$ and surface gravity $\kappa_{\rm ch}$ as, $\beta\equiv \{-\textrm{min}~(\textrm{Im}~\omega_{n})/\kappa_{\rm ch}\}$. For $\beta>(1/2)$, the scalar field $\Phi$ is regular at the Cauchy horizon and can be extended beyond the horizon. Hence the condition $\beta>(1/2)$ signifies, whether the \scc\ is respected in the spacetime or not. As emphasized above, the addition of higher curvature terms to the field equation lead to higher regularity requirement for the metric, in order to have weak solution near the \ch\ \cite{Destounis:2019omd}. Therefore, extendibility of metric perturbation beyond the \ch\ would yield a different bound on $\beta$. However, since in our analysis we consider only the case of scalar perturbation, the above bound on $\beta$ still ensures the violation of strong cosmic censorship conjecture.

Another technical point must be emphasized here, for the $\beta>(1/2)$ condition to have any relevance with the violation of the \scc, it is necessary that the late time decay of the perturbations is exponential. This is certainly true for asymptotically de Sitter black holes in general relativity \cite{Hintz:2015jkj}, but whether such an exponential decay holds for asymptotically de Sitter black holes in higher curvature theory as well, must be properly addressed. This is important, since the late time exponential decay of the perturbation is one of the essential ingredient in the analysis of \scc\ and thus for our results to make sense, we must establish such late time exponential tail for the asymptotically de Sitter black hole under consideration. Interestingly, such an exponential tail has already been reported for asymptotically de Sitter black holes in Einstein-Gauss-Bonnet gravity in \cite{Abdalla:2005hu}. Since the black hole solutions we will study in this work are also asymptotically de Sitter black holes in the Einstein-Gauss-Bonnet gravity, thus following \cite{Abdalla:2005hu}, we can safely argue that $\beta>(1/2)$ will also characterize the violation of \scc\ in the present situation. Since the properties of Gauss-Bonnet gravity, which is the second order term in the Lovelock polynomial, closely matches with the higher order terms in the full Lovelock polynomial, it is reasonable to expect that the same exponential tail would appear even for the asymptotically de Sitter black holes in pure Lovelock theories, which are also considered in this work. It is certainly possible to support the argument by further numerical analysis, e.g., time evolution of the perturbation. However, given the complicated nature of the Lovelock polynomial and the associated field equations, such an analysis is beyond the scope of this work.

In the subsequent sections we have carried out the analysis presented above in the context of a scalar field living on the charged Einstein-Gauss-Bonnet-de Sitter black hole background and subsequently for a pure lovelock black hole background. The strategy we follow here is identical to \cite{PhysRevLett.120.031103}, i.e., we start by computing the \qnm mode frequencies associated with the \PS modes, \DS modes and \NE modes numerically. Having determined each of these modes individually, we look for any possible region of parameter space for which violation of \scc\ occurs, i.e., the parameter $\beta$ becomes greater than $(1/2)$. Since the \qnm mode spectrum for \EGB as well as pure Lovelock black holes are different from those in Einstein's gravity and strongly depends on the Gauss-Bonnet coupling constant \cite{Abdalla:2005hu,Konoplya:2004xx,Konoplya:2008ix,Cuyubamba:2016cug,Konoplya:2017ymp,Konoplya:2017lhs,Nojiri:2001aj,Nojiri:2001ae}, it is reasonable to expect that the fate of strong cosmic censorship conjecture in such theories would be different and hence a detailed analysis in this context is very important. For numerical computation, we follow the Mathematica package developed in \cite{Jansen:2017oag}. Since it is expected that the \EH action must be supplemented by higher curvature terms, it is reasonable for one to expect that problems like violation of \scc\ should be settled in such higher curvature theories. This is what we explore next.
\section{Strong cosmic censorship conjecture in Einstein-Gauss-Bonnet gravity}\label{Section 3}

The statement of \scc, i.e., the assertion that solutions of Einstein's equations are non-extendible beyond Cauchy horizon, has been tested for numerous black hole solutions, but mostly within the realm of \gr. Even though certain non-trivial matter couplings are taken into account, influence of higher curvature terms on \scc\ have not been studied earlier. Since \gr\ is not a complete theory of gravity, it is crucial to understand the effects of these higher curvature modifications to \gr\ and hence on the \scc. In this work we will be interested in the higher curvature corrections within the domain of \LL Lagrangian, since they represent the most general extension to \gr\ in dimensions higher than four with field equations containing upto second derivatives of the metric. The \LL Lagrangian is a homogeneous polynomial in the Riemann tensor and is given by \cite{Lovelock:1971yv,ZWIEBACH1985315,PhysRevLett.55.2656,ZUMINO1986109,PADMANABHAN2013115,Chakraborty:2014joa},
\begin{align}
\mathcal{L} = \sqrt{-g}\sum_{k=0}^{k_{\rm max}} c_{k} L_{k}~,
\end{align}
where,
\begin{align}
L_{k}=\frac{1}{2^{k}}\delta^{a_{1}b_{1}\cdots a_{k}b_{k}}_{c_{1}d_{1}\cdots c_{k}d_{k}}R_{a_{1}b_{1}}^{c_{1}d_{1}}\cdots R_{a_{k}b_{k}}^{c_{k}d_{k}}~.
\end{align}
Here $R_{ab}^{cd}$ represents the Riemann tensor in d spacetime dimensions and $\delta^{a_{1}b_{1}\cdots a_{k}b_{k}}_{c_{1}d_{1}\cdots c_{k}d_{k}}$ denotes the totally antisymmetric Kronecker delta. The zeroth order ($k=0$) term of the \LL polynomial is the cosmological constant and the first order term ($k=1$) represents the Einstein-Hilbert Lagrangian and the second order term ($k=2$) is the Gauss-Bonnet Lagrangian. Further, $k_{\rm max}$ appearing in the \LL Lagrangian is related to the spacetime dimensions as $2k_{\rm max}\leq d$. The action for such a theory involving the first three non-trivial contributions to the \LL Lagrangian is of the following form, 
\begin{equation}
\mathcal{A}=\frac{1}{16\pi}\int d^{d}x~\sqrt{-g}\left[R+ \alpha\left(R^{2}-4R_{ab}R^{ab}+R_{abcd}R^{abcd}\right)-2\Lambda -4\pi F_{pq}F^{pq} \right]~,
\end{equation}
where we have included a matter Lagrangian of the form $-(1/4)F_{ab}F^{ab}$ and $\alpha$ is the Gauss-Bonnet coupling parameter. There exists a spherically symmetric and static black hole solution in $d$ spacetime dimensions arising out of the above action, with the line element in the form presented in \ref{generalsphsymm}, where the function $f(r)$ becomes \cite{Guo:2018exx,Cai:2001dz,Cvetic:2001bk,Hendi:2015pda},
\begin{equation}
f(r) = 1+ \frac{r^2}{2\tilde{\alpha}}\left[1-\sqrt{1+\frac{64\pi \tilde{\alpha} M}{(d-2)\Sigma_{d-2}r^{d-1}} - \frac{2\tilde{\alpha} Q^2}{(d-2)(d-3)r^{2d-4}}+\frac{8\tilde{\alpha}\Lambda}{(d-1)(d-2)}}\right]~.
\end{equation} 
Here `$Q$' is the electromagnetic charge corresponding to the field tensor $F_{\mu\nu}$ and $\tilde{\alpha}=(d-3)(d-4)\alpha$ is the rescaled Gauss-Bonnet coupling constant and $M$ is the mass of the black hole. Further, $\Sigma_{d-2}$ is the volume of a $(d-2)$ dimensional unit sphere. The location of the horizons are given by the equation $f(r)=0$, which further reduces to,
\begin{align}\label{horizon}
\frac{4\,\Lambda}{(d-1)(d-2)}r^{(2d-4)}-2\,r^{(2d-6)}-2\,\alpha\, r^{(2d-8)}+\frac{32\pi M}{(d-2)\Sigma_{d-2}}\, r^{(d-3)}- \frac{Q^2}{(d-2)(d-3)}=0
\end{align}
Since, for our analysis we require the black hole under consideration to have three horizon, namely the event horizon, \coh and \ch, \ref{horizon} must gives rise to three real positive root. This is guaranteed from the Descarte rule of sign. 

Given this black hole spacetime, which is an exact solution of the higher curvature gravitational field equations, we are interested in studying if there is any violation of \scc\ in this spacetime. It should be emphasized that when $\alpha=0$, i.e., in the absence of higher curvature terms, the spacetime reduces to a \RN\DS configuration in $d-$dimensions and admits violation of \scc \cite{Rahman:2019uwf}. It is therefore interesting to see whether the addition of higher curvature terms may cure the violation of \scc\ when $\alpha \neq 0$. Given the choice for $f(r)$, one can explicitly determine the quantity $\beta$ for the photon sphere modes by computing the Lyapunov exponent and surface gravity at the Cauchy horizon following \ref{generalbeta}. However, as emphasized earlier in this section, it is better to determine the \qnm mode frequencies numerically and then further obtain $\beta$ to demonstrate the violation of \scc.

Let us start by describing the dynamics of a massless scalar field $\Phi$ on a $d$ dimensional spherically symmetric black hole background as given in \ref{generalsphsymm}. The evolution of the perturbation is governed by the Klein-Gordon equation $\square \Phi=0$. For spherically symmetric background, one can always expand the field in terms of a natural basis on the $(d-2)$ sphere, namely the spherical harmonics $\mathcal{Y}_{l\,m}(\theta,\phi)$ as follows,
\begin{align}
\Phi(t,r,\Omega) = \sum_{\ell,m}e^{-i\omega t}\frac{\phi(r)}{r^{(d-2)/2}}\mathcal{Y}_{\ell m}(\Omega)
\end{align}
which leads to the following master equation,
\begin{align}\label{mastereqn}
\left(\frac{\partial^2}{\partial r_{*}^2}+\omega^2-V_{\rm eff}(r)\right)\phi(r)=0;\quad 
V_{\rm eff}(r) = f(r)\left\{\frac{\ell(\ell+d-3)}{r^2}+\frac{(d-2)(d-4)}{4r^2}f(r)+\frac{(d-2)}{2r}f'(r)\right\}
\end{align}
where, $dr_{*}=\{dr/f(r)\}$ is the tortoise coordinate and $V_{\rm eff}(r)$ is the effective potential expressed above in terms of the metric function $f(r)$. The \qnm mode frequency $\omega_{n}$ is defined as the eigenvalue of \ref{mastereqn} that corresponds to ingoing modes at the event horizon, $r_{\rm h}$ and outgoing modes at the cosmological horizon, $r_{\rm c}$, i.e.,
\begin{align}
\phi(r\rightarrow r_h)\sim e^{-i\omega r_*}\qquad \text{and} \qquad \phi(r\rightarrow r_c)\sim e^{i\omega r_*}
\end{align}
As mentioned earlier, for computing the quasi-normal modes numerically we follow the procedure and use the Mathematica package developed in \cite{Jansen:2017oag}. This requires one to work with the redefined radial coordinate $u=1/r$ and impose the quasi-normal mode boundary conditions appropriately at the event horizon and at the cosmological horizon. Thus we obtain the complex frequencies of the \qnm modes, which is the first ingredient that goes into the definition of $\beta$. The computation of $\kappa_{\rm ch}$ can also be performed in a similar manner and hence the numerical estimation for $\{-(\textrm{Im}~\omega_{n,\ell})/\kappa_{\rm ch}\}$ can be obtained, whose minimum value would yield an estimation for $\beta$. This has been shown explicitly in \ref{table-QNM-GB} for the three modes of interest, namely the \NE modes ($\ell=0$), the \DS modes ($\ell=1$) and the \PS modes ($\ell=10$). Analytical estimations for $\{-(\textrm{Im}~\omega_{n,\ell})/\kappa_{\rm ch}\}$ has also been presented in \ref{table-QNM-GB} and as evident from the numerical results, the analytical and numerical values matches quite well, within an error of 6\%. Further, from \ref{table-QNM-GB} we see that $\beta$ crosses the value $(1/2)$ for near extremal values of the charge parameter $Q$ and hence the violation of \scc\ does occur in the context of asymptotically de Sitter black holes in Einstein-Gauss-Bonnet gravity inheriting Cauchy Horizon.

\begin{table}[h]
\begin{centering}
\begin{tabular}{|c|c|c|c|c|c|c|c|}
\hline 
$\alpha$ & $\Lambda$ & $Q/Q_{\textrm{max}}$ & $\ell=0$ & $\ell=1$ & $\ell=10$ & $\ell=10$ (analytical)\tabularnewline
\hline 
\hline 
\multirow{4}{*}{$0.1$} & \multirow{2}{*}{$0.06$} & \multirow{1}{*}{$0.99$}  & $0.849266$ & $0.467428 $ & $0.678101 $& $0.6770764 $\tabularnewline

\cline{3-7} 
 &  & \multirow{1}{*}{$0.995$} & $0.8860955$ &  $0.7059601$ & $1.02414 $& $1.018981$\tabularnewline

\cline{2-7} \cline{3-7}  
 & \multirow{2}{*}{$0.1$} & \multirow{1}{*}{$0.99$} & $0.850344$ &  $0.6296578 $ & $0.6674398 $& $0.6661463 $\tabularnewline
 
\cline{3-7}
 &  & \multirow{1}{*}{$0.995$}  & $0.8841346$ & $0.9521732 $ & $1.00561$& $1.00365144 $\tabularnewline
 
\cline{7-7}  
\hline 
\multirow{4}{*}{$0.2$} & \multirow{2}{*}{$0.06$} & \multirow{1}{*}{$0.99$}  & $0.861229$ & $0.510842 $ & $0.734559 $& $0.7334486$\tabularnewline

\cline{3-7} 
 &  & \multirow{1}{*}{$0.995$} & $0.8952727$ &  $0.7683285 $ & $1.10481 $& $1.09940317$\tabularnewline

\cline{2-7} \cline{3-7}  
 & \multirow{2}{*}{$0.1$} & \multirow{1}{*}{$0.99$} & $0.8608527$ &  $0.685808 $ & $0.7215474 $& $0.7201545 $\tabularnewline
 
\cline{3-7}
 &  & \multirow{1}{*}{$0.995$}  & $0.893013$ & $1.0327756 $ & $1.0865954 $& $1.0806729 $\tabularnewline
 
\cline{7-7} 
\hline 
\multirow{4}{*}{$0.3$} & \multirow{2}{*}{$0.06$} & \multirow{1}{*}{$0.99$}  & $0.8714403$ & $0.555417 $ & $0.796272 $& $0.79027861 $\tabularnewline

\cline{3-7} 
 &  & \multirow{1}{*}{$0.995$} & $0.9035516$ &  $0.83237279 $ & $1.18615 $& $1.1804926 $\tabularnewline

\cline{2-7} \cline{3-7}  
 & \multirow{2}{*}{$0.1$} & \multirow{1}{*}{$0.99$} & $0.8703119$ &  $0.743391 $ & $0.776177 $& $0.774676 $\tabularnewline
 
\cline{3-7}
 &  & \multirow{1}{*}{$0.995$}  & $0.9010936$ & $1.1544405 $ & $1.160704 $& $1.1584423 $\tabularnewline
\cline{7-7} 
\hline 
\end{tabular}
\par\end{centering}
\caption{Numerical values of $\{-(\textrm{Im}~\omega_{n,\ell})/\kappa_{\rm ch}\}$ have been presented for the lowest lying quasi-normal modes for different choices of $\ell$. We have also presented them for various choices of the rescaled Gauss-Bonnet coupling constant $\tilde{\alpha}$, cosmological constant $\Lambda$ and rescaled electric charge $(Q/Q_{\rm max})$, for $M=1$ and $d=5$. The numerical estimation of $\beta$, for a given $\tilde{\alpha}$, $\Lambda$ and $(Q/Q_{\rm max})$ would correspond to the lowest entry in that respective row. The values presented in the first column, with $\ell=0$, correspond to the near-extremal modes, while the second column, with $\ell=1$, depict the de Sitter modes. Finally, the numerical estimation for $\{-(\textrm{Im}~\omega_{n,\ell})/\kappa_{\rm ch}\}$ associated with the \PS modes have been presented for $\ell=10$. To see the direct correspondence with the analytical results presented in \ref{Section 2}, in the last column we provide the analytical estimate of the same as well. As evident the numerical and analytical estimations of $\{-(\textrm{Im}~\omega_{n,\ell})/\kappa_{\rm ch}\}$ are in close agreement, thereby justifying the use of analytical techniques for black holes in higher curvature theories of gravity.}
\label{table-QNM-GB}
\end{table}

Let us now verify the violation of the \scc\ in an explicit manner. This can be achieved by plotting $\{-(\textrm{Im}~\omega_{n,\ell})/\kappa_{\rm ch}\}$ (which for brevity have been labelled as $\beta$) with respect to $(Q/Q_{\rm max})$, where $Q_{\rm max}$ is the extremal limit of the electric charge $Q$ for a given cosmological constant and Gauss-Bonnet parameter $\alpha$, in \ref{5d_GB}. The left column of the figure depicts the \PS modes, the plots in the middle column depicts the \DS modes and finally the plots on the right-most column illustrates the \NE modes. It is again obvious that all of these modes crosses $\beta=(1/2)$ line and hence \scc\ is violated for charged, asymptotically de Sitter black holes in Einstein-Gauss-Bonnet gravity. Thus presence of higher curvature terms do \emph{not} help to restore \scc. Furthermore, the violation gets severe as the Gauss-Bonnet coupling parameter $\alpha$ is increased, since the curves for $\beta$ crosses the line $\beta=(1/2)$ earlier, thus allowing for a larger parameter space where the violation of \scc\ can be perceived. Further, for \PS modes the violation becomes stronger as the spacetime dimension is increased from $d=5$ to $d=6$ (see, the last row of \ref{5d_GB}), which is a reminiscent of the result presented in \cite{Rahman:2018oso}.   

\begin{figure}[!htp]
\centering
\subfloat{{\includegraphics[scale=0.37]{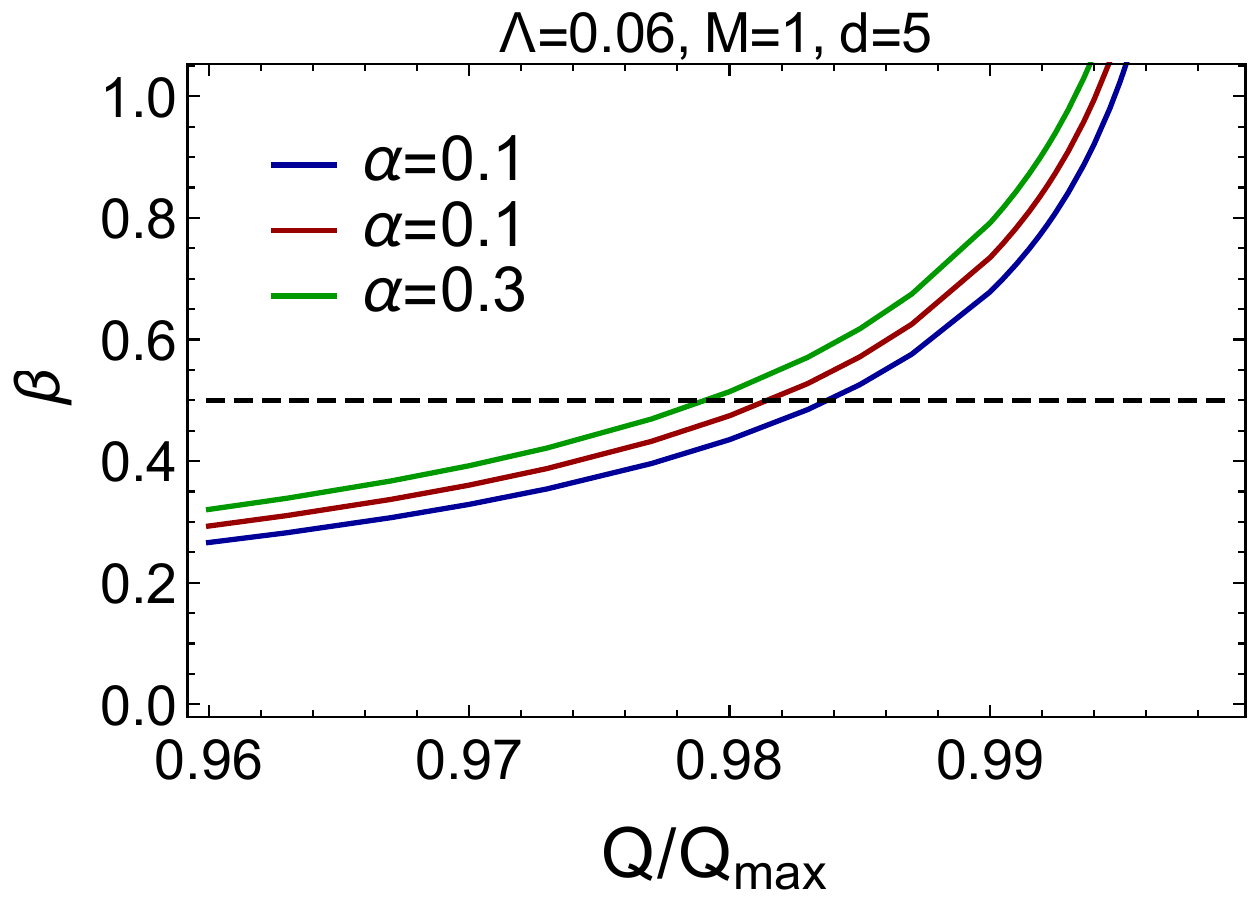} }}    
\qquad
\subfloat{{\includegraphics[scale=0.37]{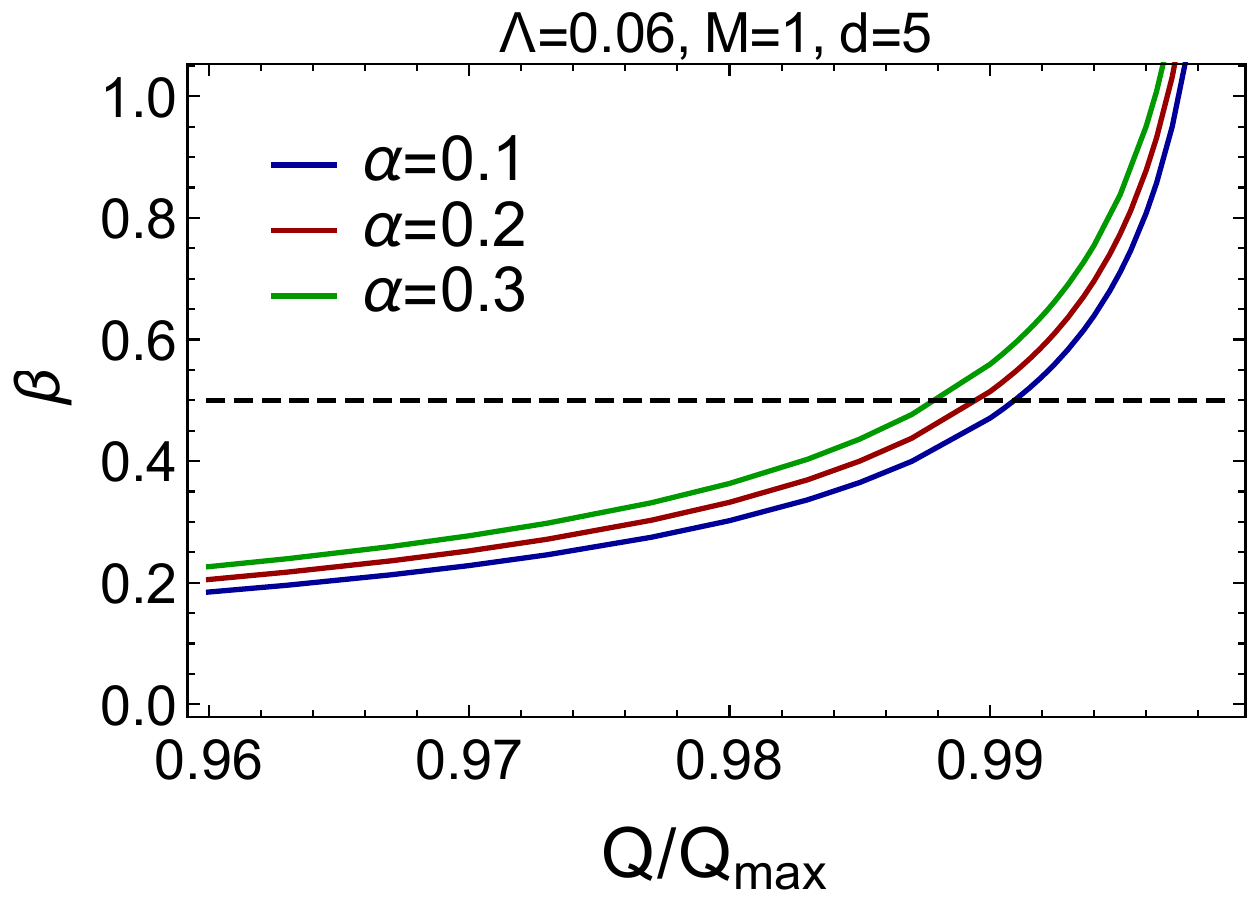} }}
\qquad
\subfloat{{\includegraphics[scale=0.39]{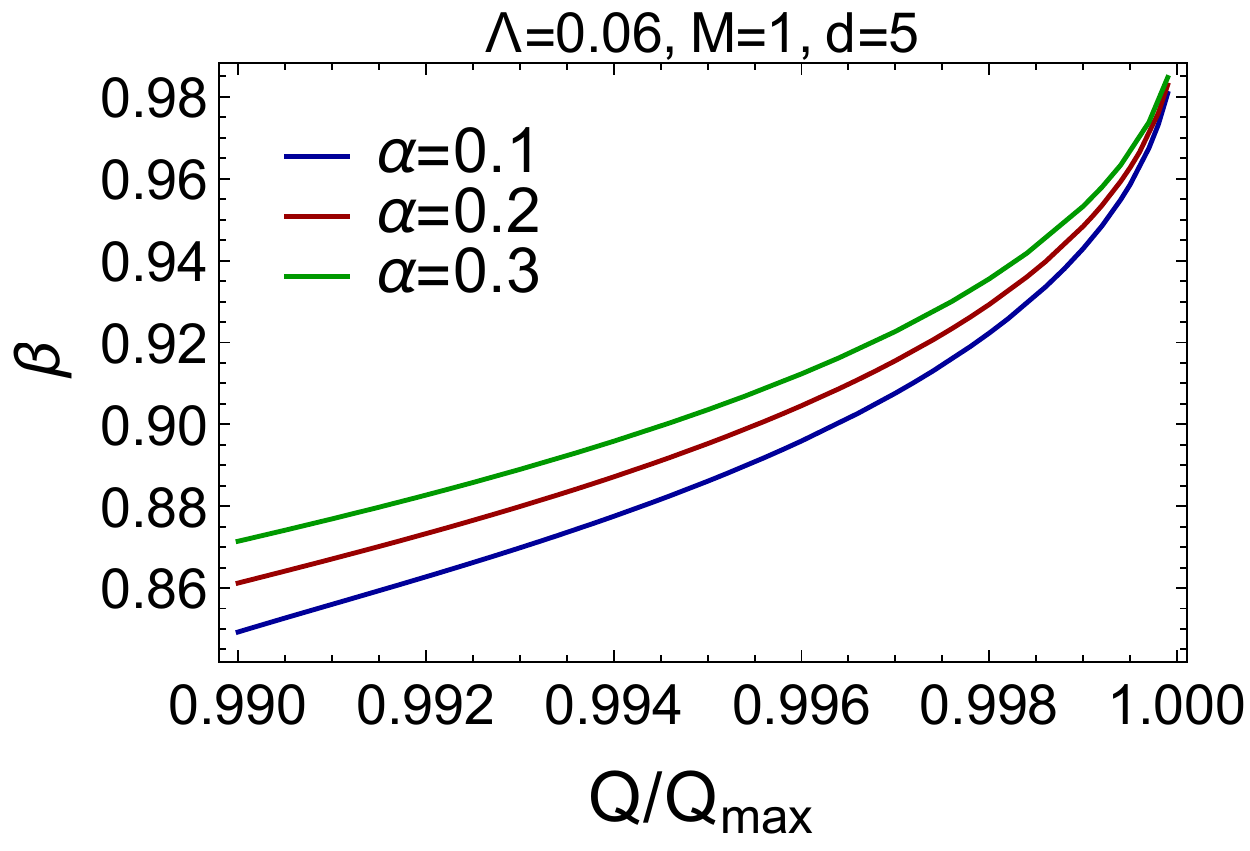} }}
\qquad
\subfloat{{\includegraphics[scale=0.37]{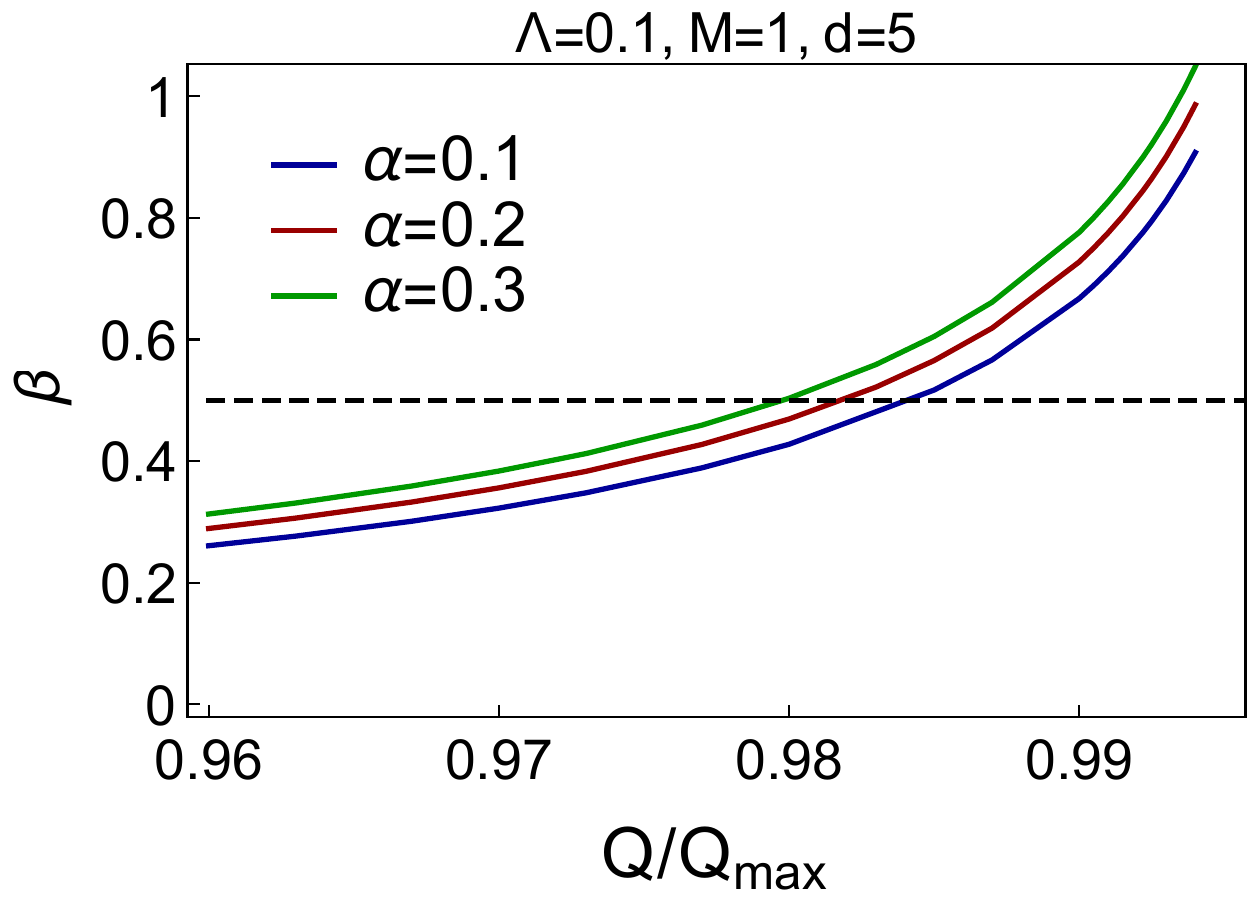} }}
\qquad
\subfloat{{\includegraphics[scale=0.37]{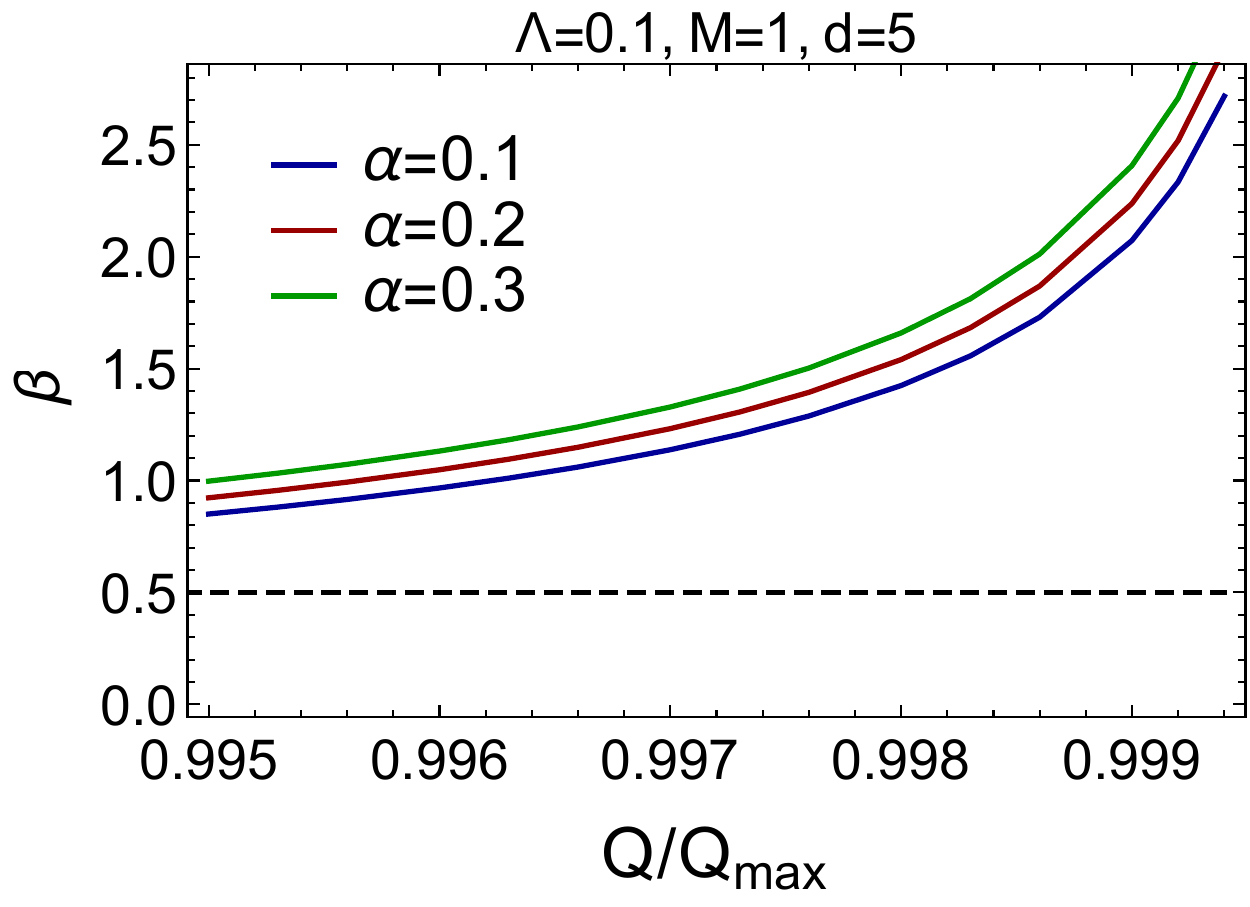} }}
\qquad
\subfloat{{\includegraphics[scale=0.39]{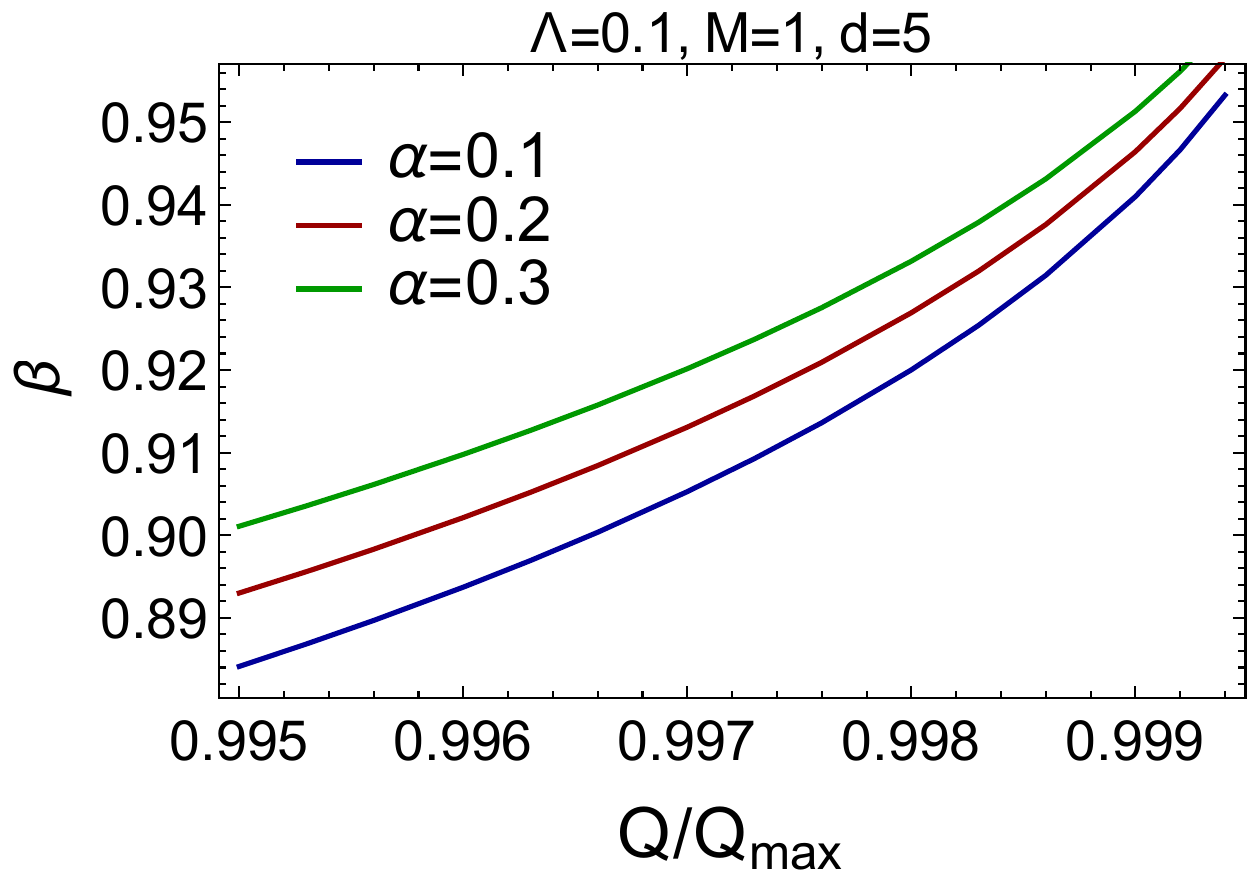} }}
\qquad
\subfloat{{\includegraphics[scale=0.37]{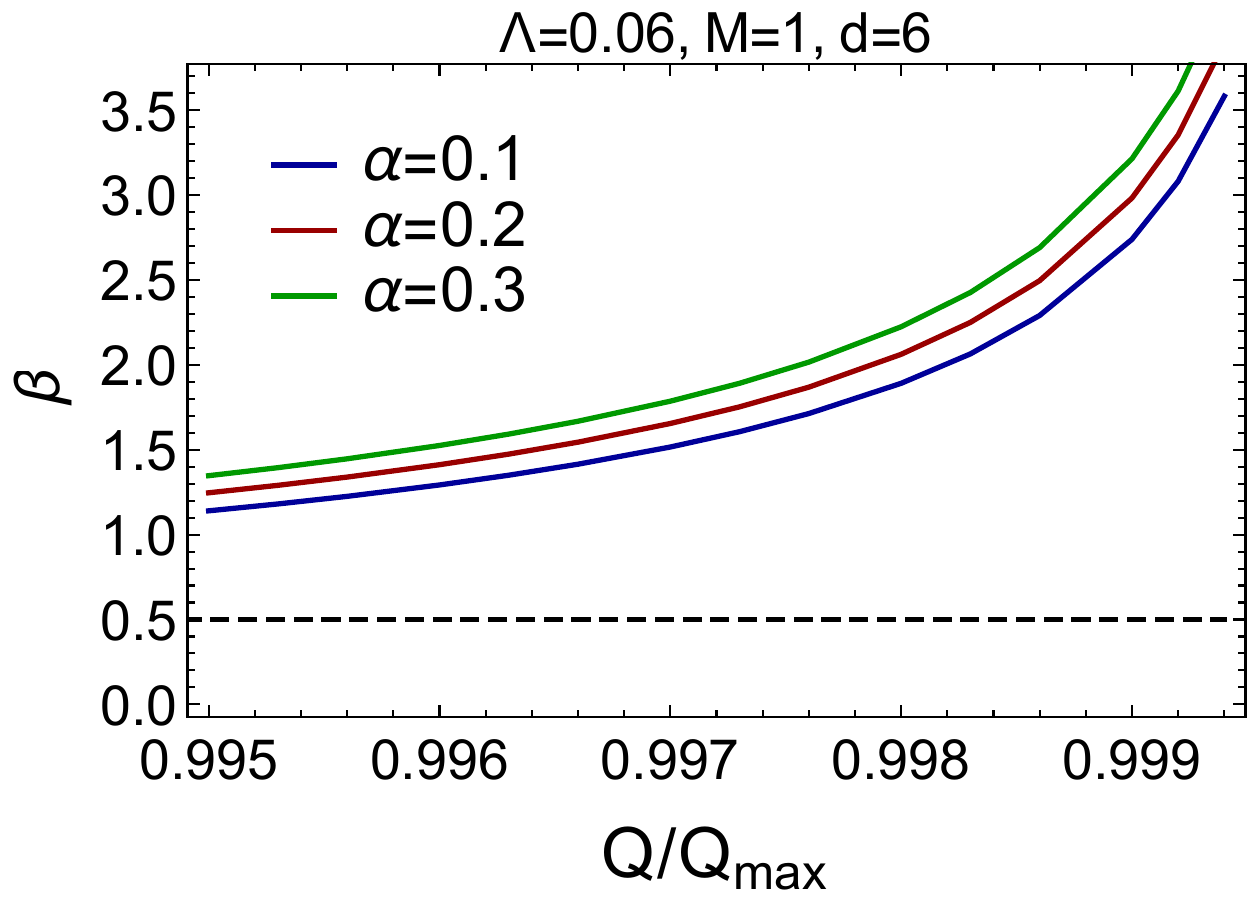} }}
\qquad
\subfloat{{\includegraphics[scale=0.37]{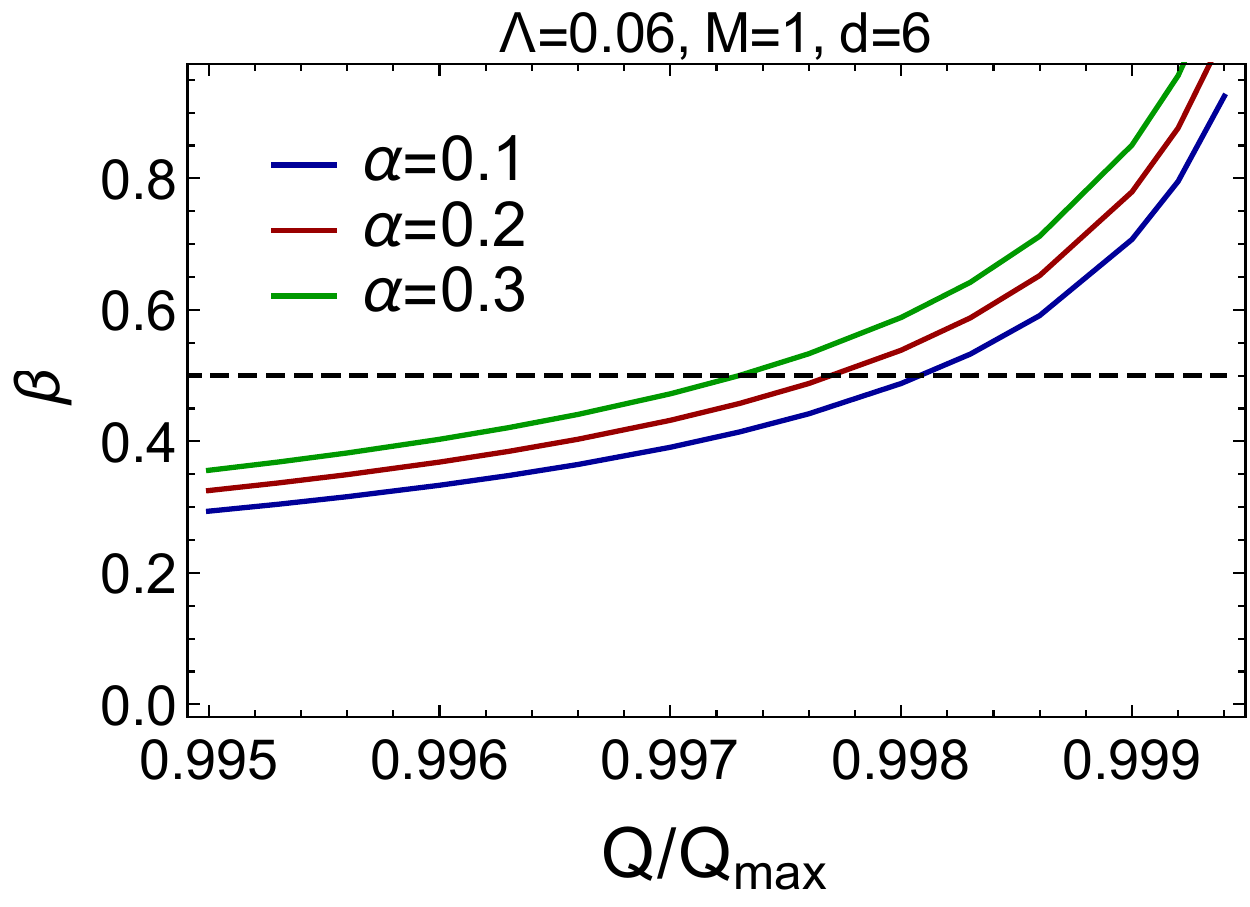} }}
\qquad
\subfloat{{\includegraphics[scale=0.39]{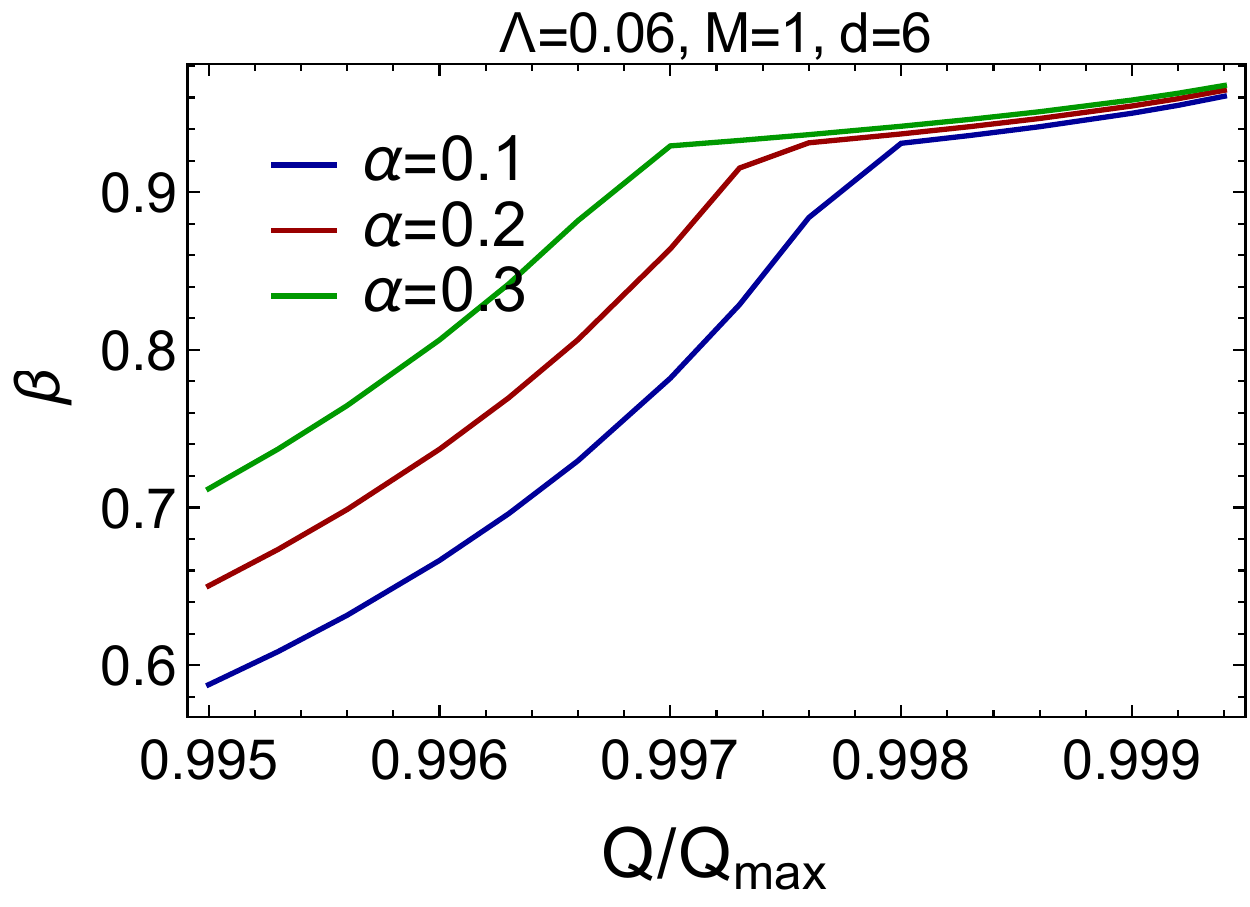} }}
\caption{We have plotted the quantity $\{-(\textrm{Im}~\omega_{n})/\kappa_{\rm ch}\}$, whose minima provides an estimation for $\beta$, with the ratio $(Q/Q_{\rm max})$ for all the three \qnm modes of different origins. The plots on the leftmost column depicts the variation of the ratio $\{-(\textrm{Im}~\omega_{n})/\kappa_{\rm ch}\}$ for the \PS modes, the plots in the middle column are for variation of the same quantity with the \DS modes and finally the plots on the right column are showing the variations with the \NE modes. All the plots in a certain row are for a fixed value of the cosmological constant $\Lambda$ and all the three curves in a given plot are for three choices of the rescaled Gauss-Bonnet parameter. See text for discussions.}
 \label{5d_GB}
\end{figure}

\begin{figure}[!htp]
\centering
\subfloat{{\includegraphics[scale=0.37]{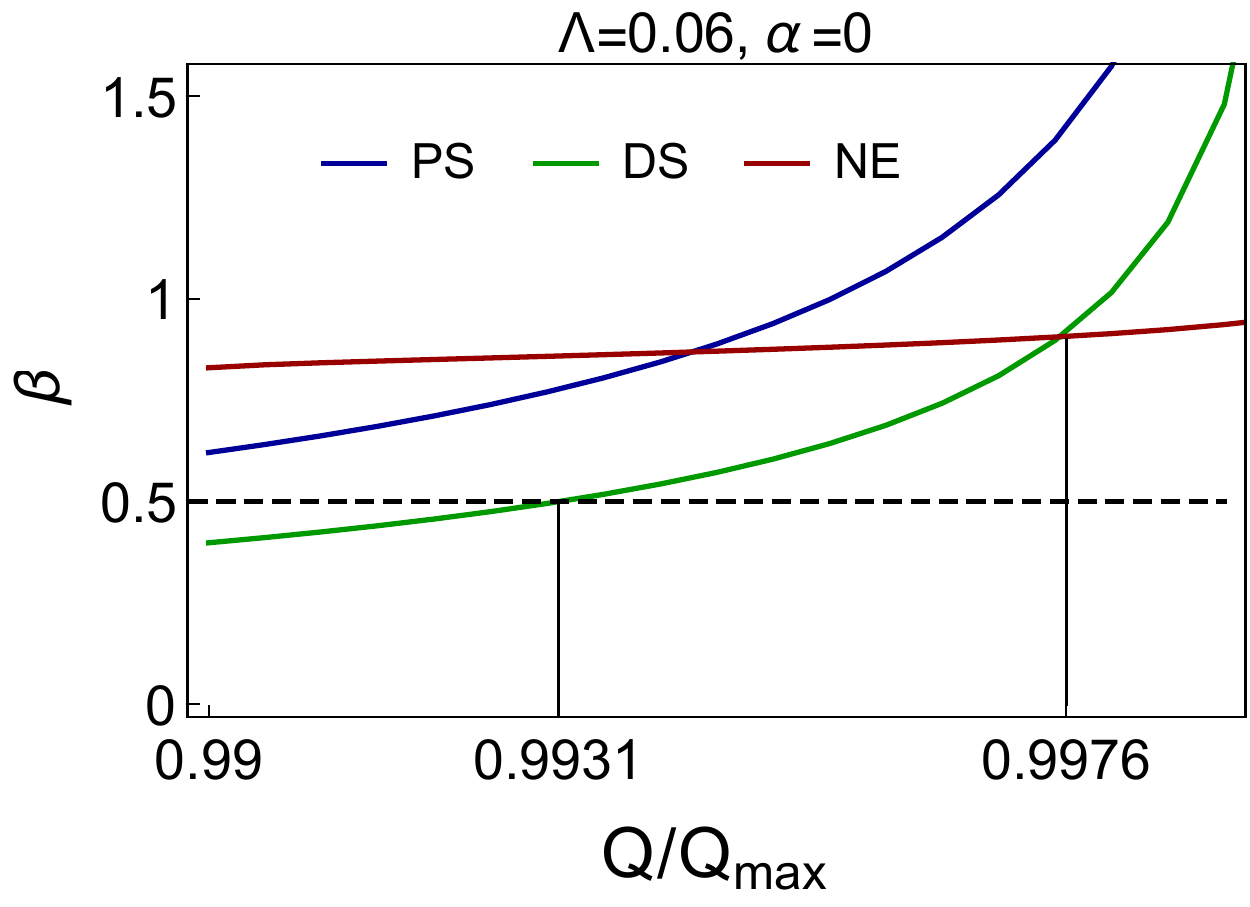} }}
\qquad
\subfloat{{\includegraphics[scale=0.37]{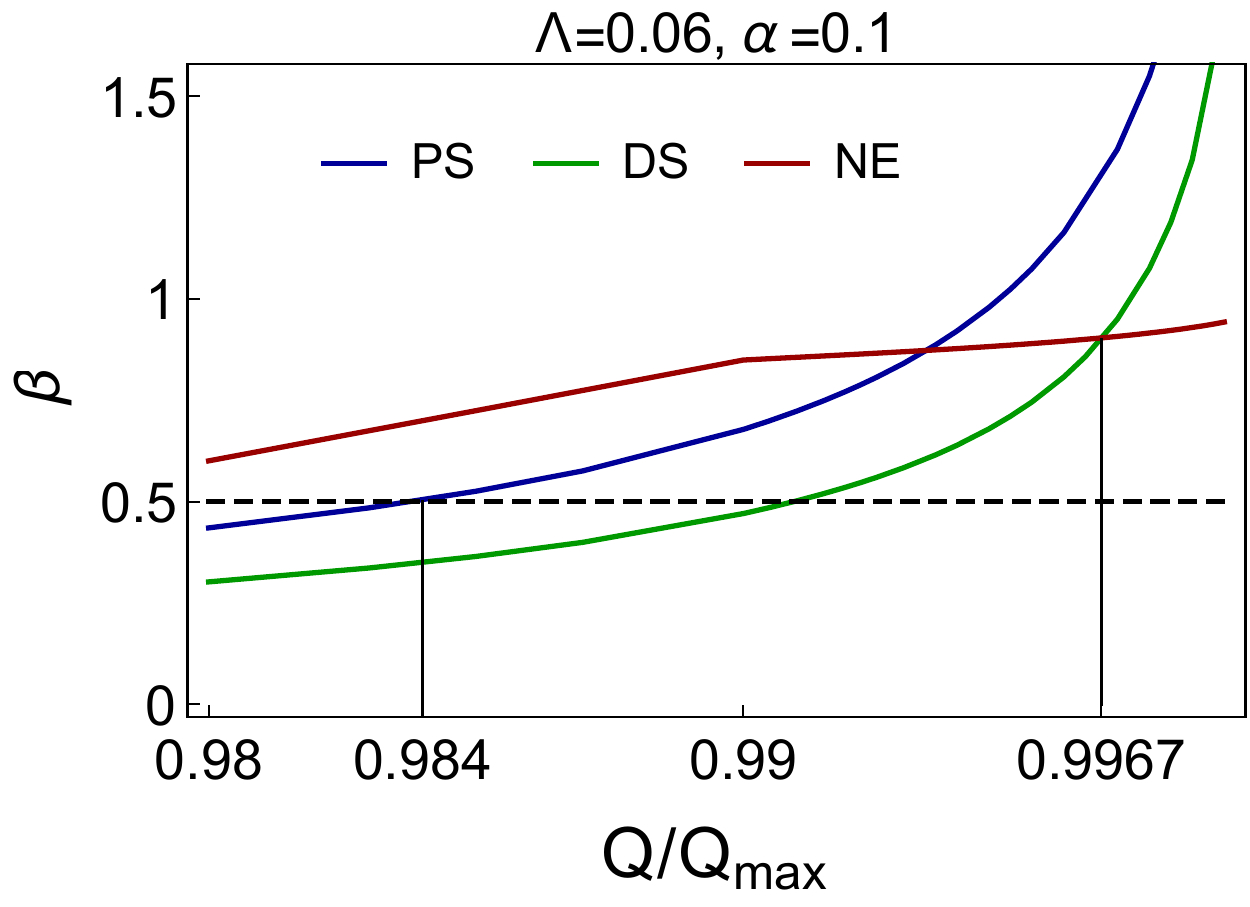} }}
\qquad
\subfloat{{\includegraphics[scale=0.37]{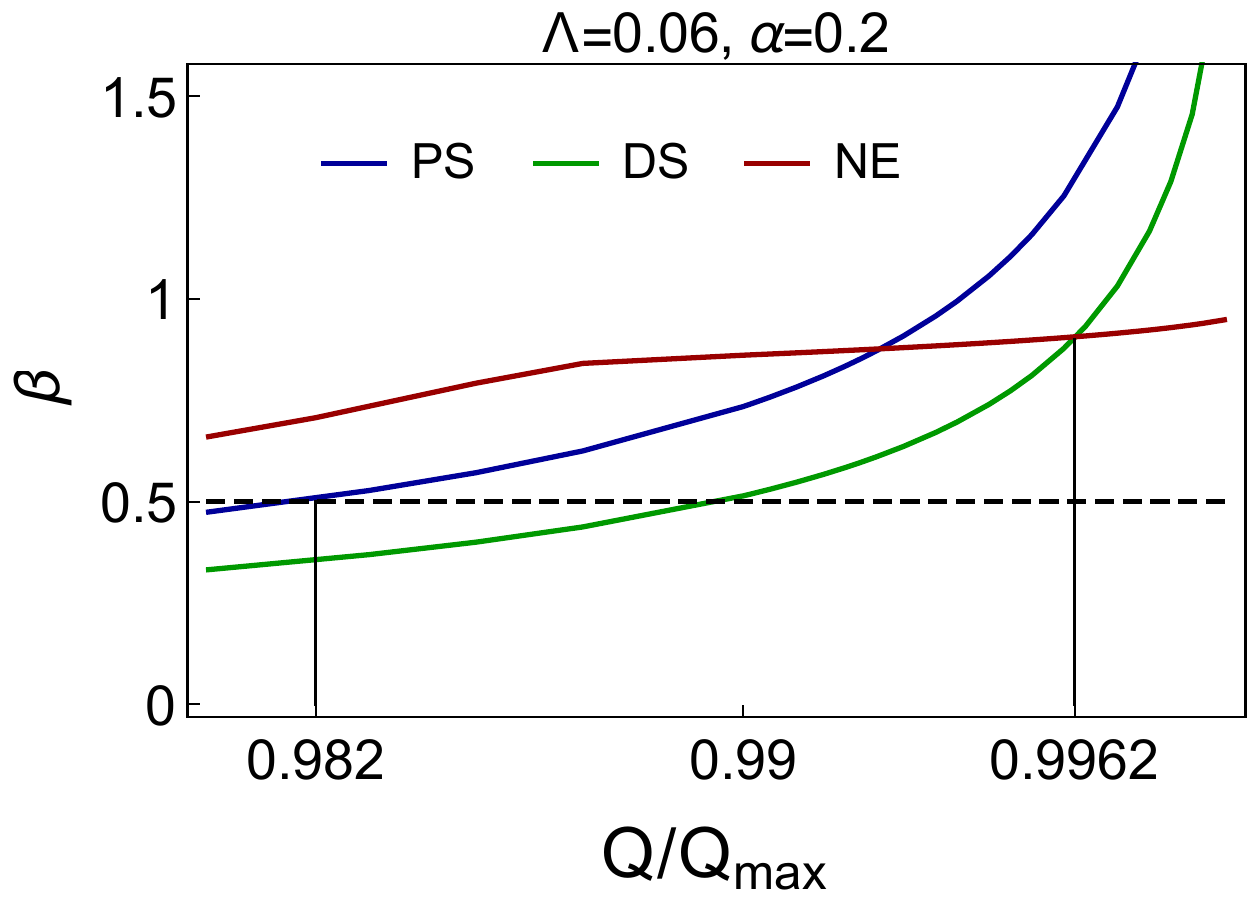} }}
\qquad
\subfloat{{\includegraphics[scale=0.37]{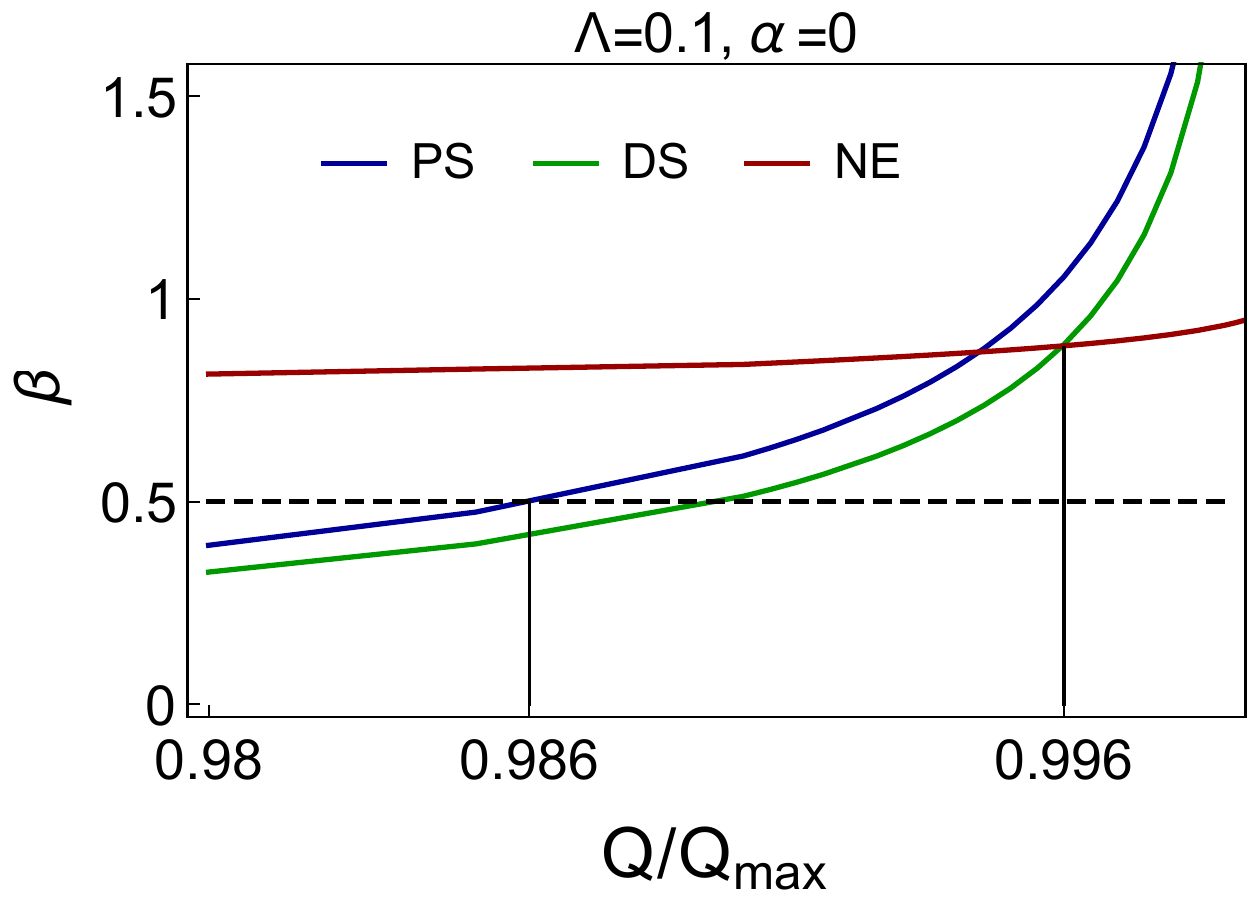} }}
\qquad
\subfloat{{\includegraphics[scale=0.37]{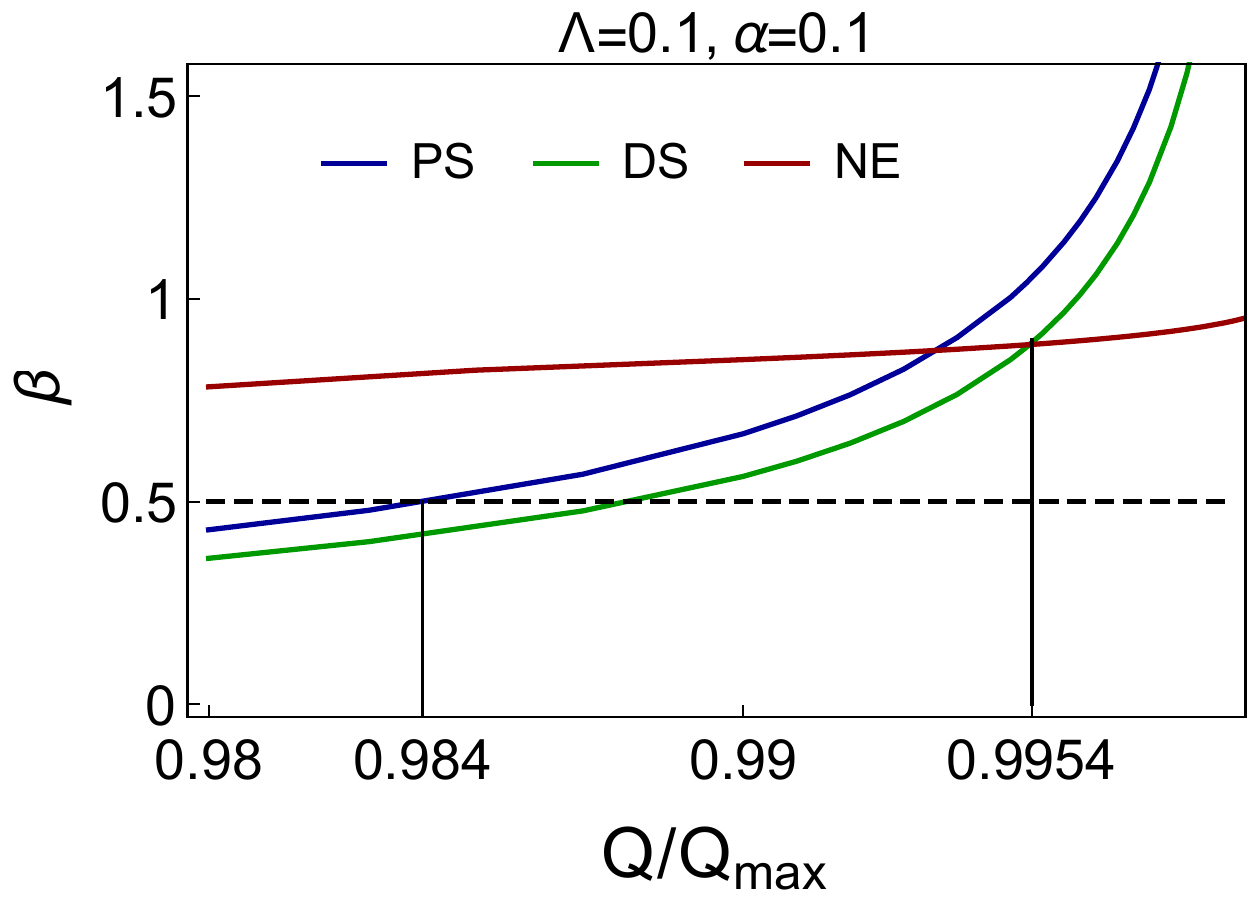} }}
\qquad
\subfloat{{\includegraphics[scale=0.37]{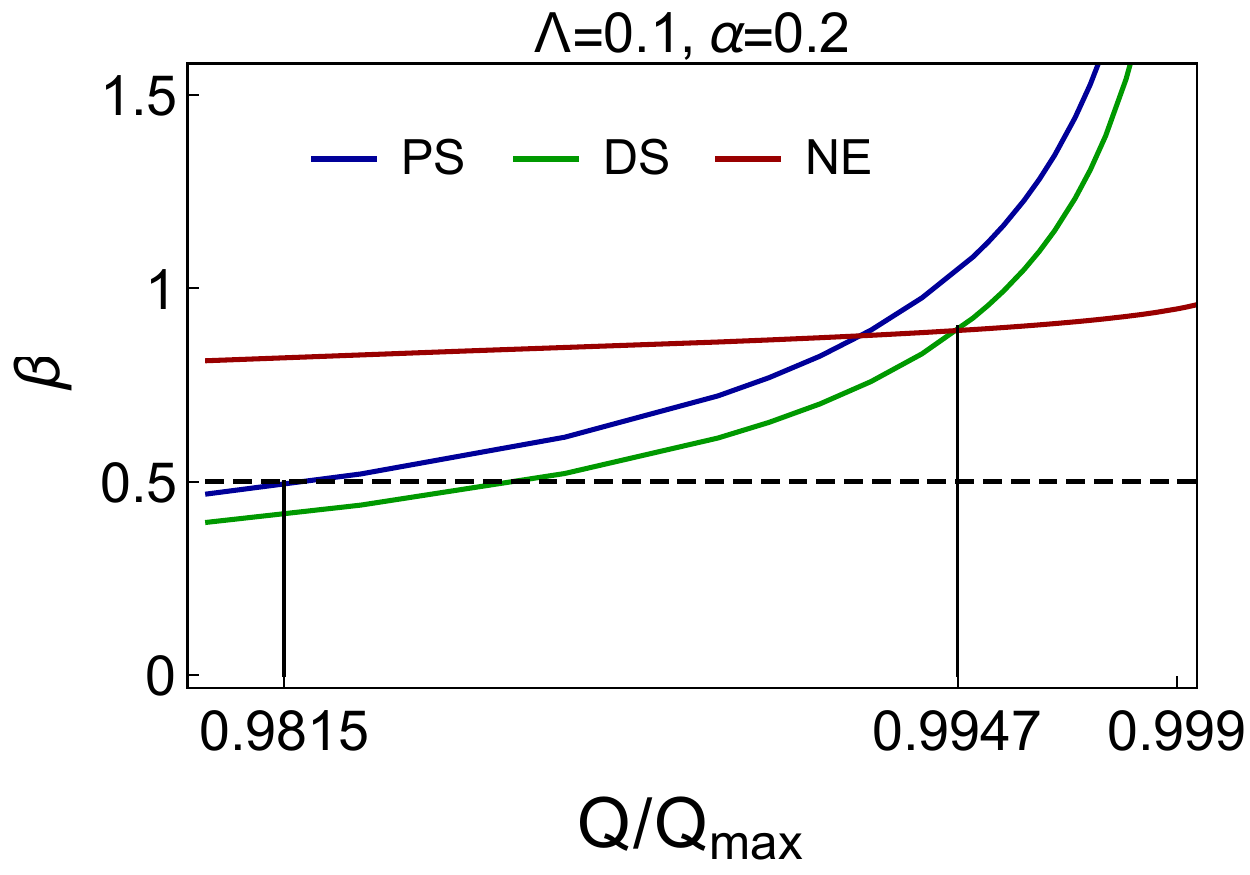} }}
\qquad
\subfloat{{\includegraphics[scale=0.37]{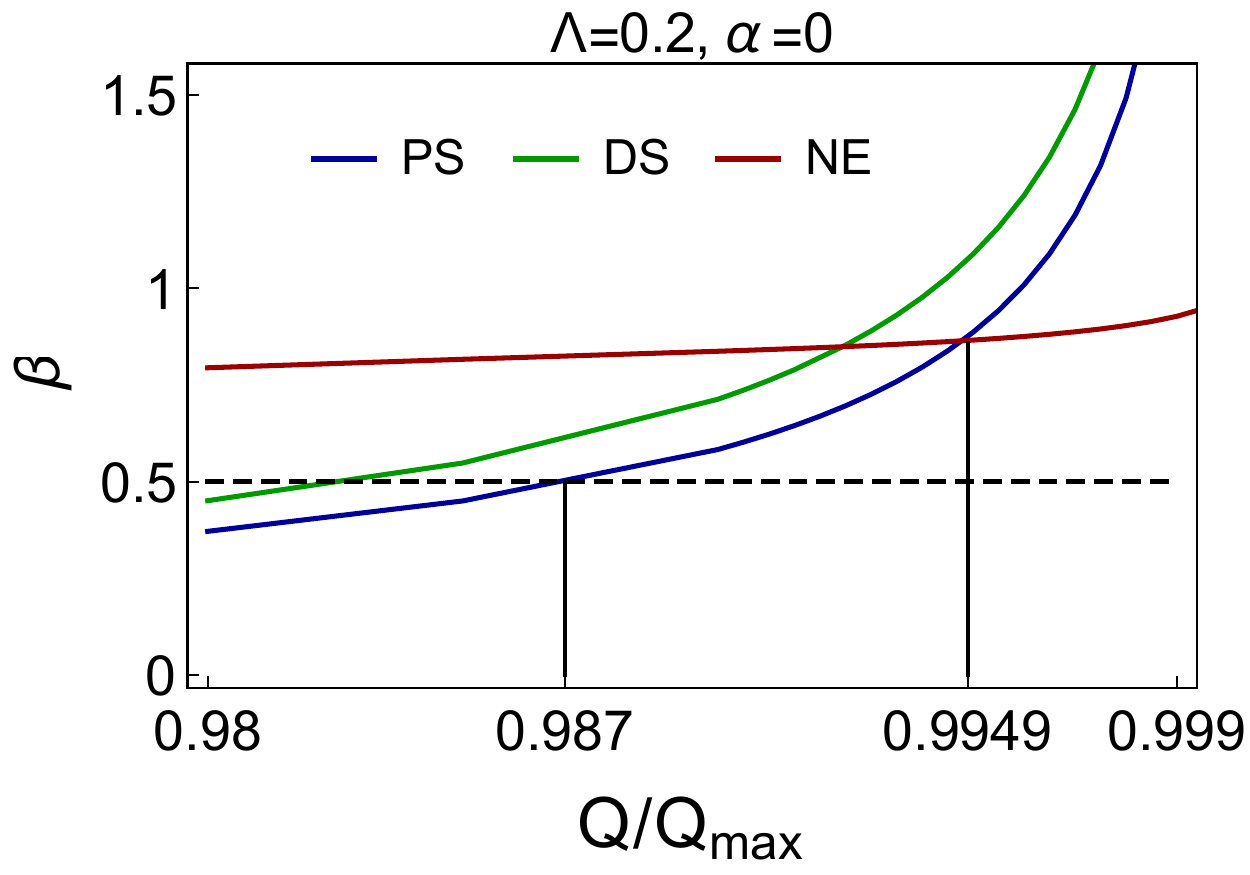} }}
\qquad
\subfloat{{\includegraphics[scale=0.37]{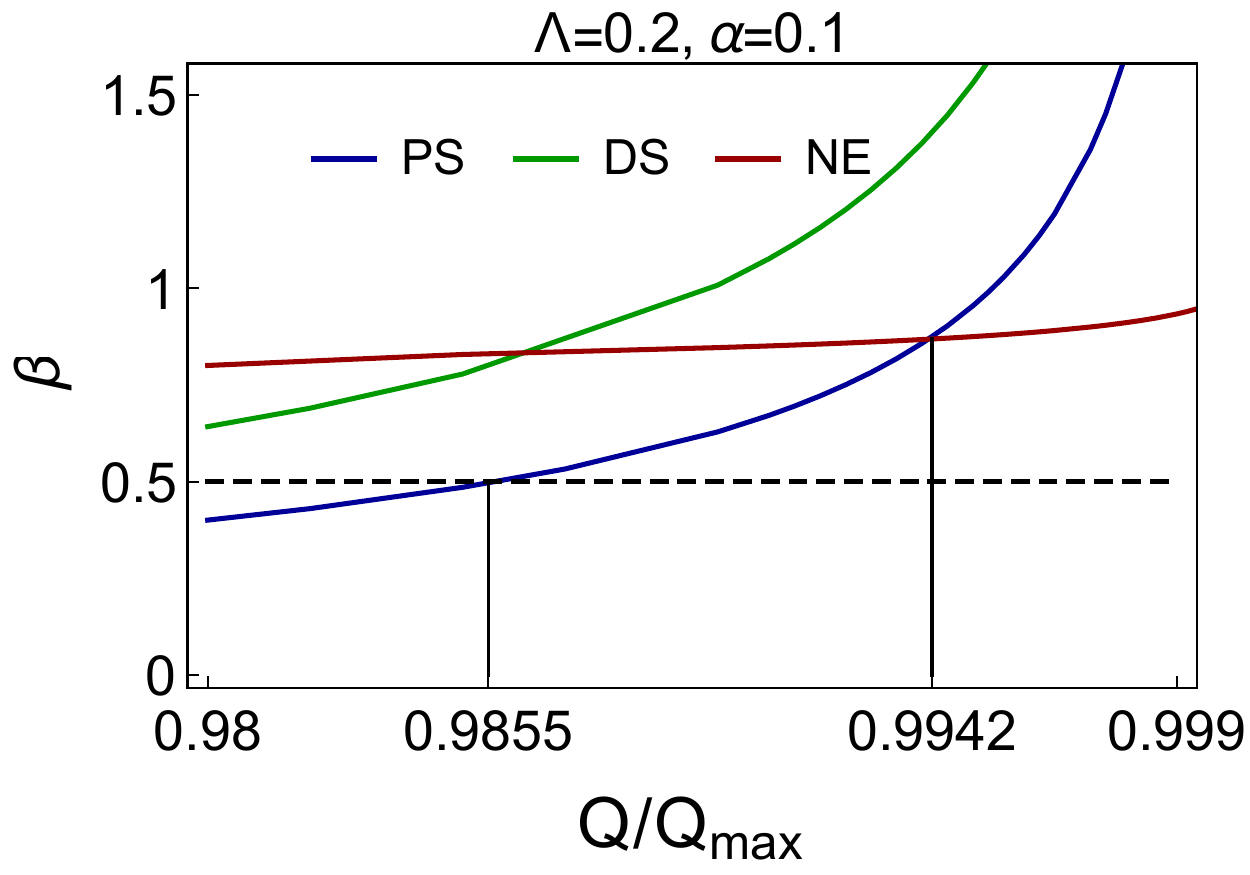} }}
\qquad
\subfloat{{\includegraphics[scale=0.37]{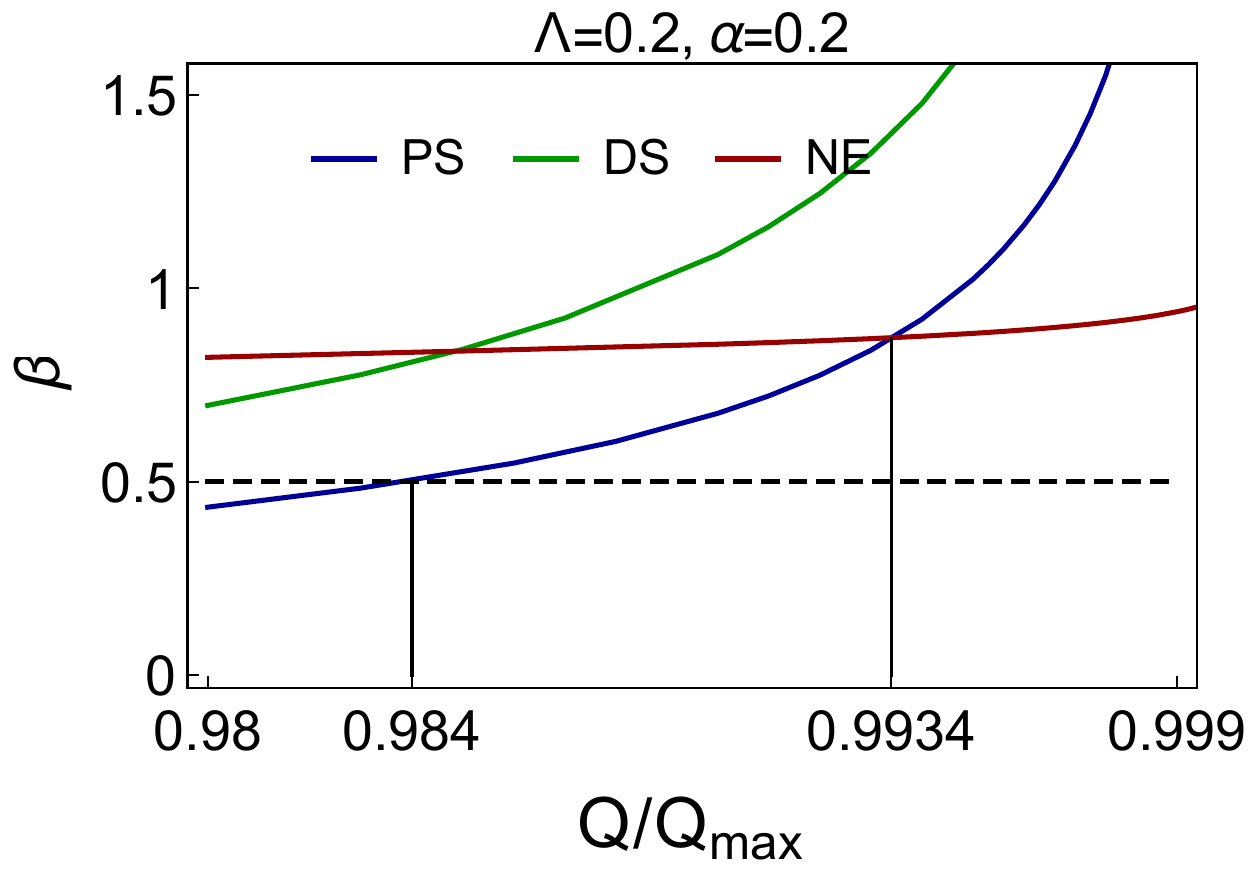} }}
\qquad
\caption{In this figure we have demonstrated the variation of $\{-(\textrm{Im}~\omega_{n})/\kappa_{\rm ch}\}$, whose minima corresponds to the parameter $\beta$, with respect to $(Q/Q_{\rm max})$ for all the three \qnm modes of interest in each single plots. The \PS modes are denoted by blue curves, the \DS modes are represented by green curves and the \NE modes are depicted by brown curves. Each of these plots are for various choices of the cosmological constant $\Lambda$ and Gauss-Bonnet parameter $\alpha$. The first vertical line in each of these plots corresponds to the value of $(Q/Q_{\rm max})$, where $\beta$ becomes greater than $(1/2)$ for the first time and hence the strong cosmic censorship conjecture is violated. While the second vertical line corresponds to the value of $(Q/Q_{\rm max})$, where the \NE modes starts to dominate. In this plot, the  mass is taken to be unity and the spacetime dimension to be $d=5$.}
 \label{GB_all_modes}
\end{figure}

To see these results from a different perspective, we have again plotted the ratio of the imaginary part of the \qnm and the surface gravity at the Cauchy horizon   against $(Q/Q_{\rm max})$, but this time with all the three modes depicted in the same plot for various choices of the cosmological constant $\Lambda$ and the Gauss-Bonnet Parameter $\alpha$ in \ref{GB_all_modes} for $d=5$. As evident from \ref{GB_all_modes}, for smaller values of $\Lambda$, the \DS mode dominates over and above the other two for smaller $(Q/Q_{\rm max})$. Subsequently, the \DS mode crosses the $\beta=(1/2)$ line, thus violating the \scc\ and finally the \NE modes take over. On the other hand, for larger values of the cosmological constant, the \PS mode dominates, which crosses the $\beta=(1/2)$ line and finally gets sub-dominant to the \NE modes. The effect of higher curvature terms on the violation of \scc\ can be easily realized from \ref{GB_all_modes} as well. Since it clearly illustrates that as the Gauss-Bonnet coupling parameter $\alpha$ increases, the violation of \scc\ happens at smaller and smaller values of $(Q/Q_{\rm max})$. This implies that the parameter space available for violating \scc\ is larger in higher curvature theories in comparison to \gr.  For this purpose, we have plotted $\beta$ against $(Q/Q_{\rm max})$ for three choices of the Gauss-Bonnet coupling parameter $\alpha$, including $\alpha=0$, which represents the general relativistic scenario.  

\section{Strong cosmic censorship conjecture in pure Lovelock gravity}\label{Section 4}

So far, our discussion has been on the violation of \scc\ in the context of five and higher dimensional Einstein-Gauss-Bonnet gravity. As illustrated in the previous section, the Einstein-Gauss-Bonnet gravity leads to even stronger violation of \scc\ than that of a black hole solution in \gr. This still kept a room for the question, what happens for other higher curvature terms in the \LL Lagrangian? In this section, we wish to study the effect of other higher curvature terms in the \LL Lagrangian on the violation of \scc\ and for this purpose we wish to consider the case of pure Lovelock gravity \cite{PhysRevD.100.084011,Dadhich:2015nua}, which refers to a single term in the full Lovelock polynomial. More precisely, $k^{\rm th}$ order pure Lovelock term corresponds to the Lagrangian $\mathcal{L}=\sqrt{-g}L_{k}$, without the sum. For example, the second-order pure Lovelock theory has the Lagrangian of the form, $\mathcal{L}_{2}=\sqrt{-g}\{R^{2}-4R_{ab}R^{ab}+R_{abcd}R^{abcd}\}$. Thus the action for such a theory involving a pure Lovelock Lagrangian with a positive cosmological constant term is given by,
\begin{equation}
\mathcal{A}=\frac{1}{16\pi}\int d^dx\!\sqrt{-g}\left[-2\Lambda+L_{k}-4\pi F_{ab}F^{ab}\right]
\end{equation}
Such a theory admits spherically symmetric black hole solution in d spacetime dimensions with the line element expressed in the form of \ref{generalsphsymm}, where the function $f(r)$ is given by \cite{Chakraborty},
\begin{align}
f(r)= 1-\left(\tilde{\Lambda}r^{2k} + \frac{2M^k}{r^{d-2k-1}}-\frac{\tilde{Q}^2}{r^{2d-2k-4}}\right)^{\frac{1}{k}}
\end{align}
where, $\tilde{\Lambda}$ and $\tilde{Q}$ are some rescaled version of the cosmological constant and $U(1)$ electromagnetic charge associated with the Maxwell field coupled with gravity. It is well known that, such a black hole solution admits instabilities with respect to small perturbations in dimensions $d<(3k+1)$ \cite{PhysRevD.100.084011}. Hence, in our analysis, we restrict our attention only to black holes in $d\geq (3k+1)$. In particular, we demonstrate our result for the pure Gauss-Bonnet solution ($k=2$) in seven spacetime dimensions. To illustrate the effect of pure Gauss-Bonnet term on the \scc, we compare it with that of the corresponding black hole solution of Einstein gravity in seven dimensions, i.e., a \RN de Sitter solution. In $d=7$, the metric component $f(r)$ for the charged pure Gauss-Bonnet-de-Sitter and charged Einstein-de-Sitter black hole solution takes the form,
\begin{align}
f_{\rm GB}(r)= 1-\left(\tilde{\Lambda}r^{4} + \frac{2M^2}{r^{2}}-\frac{\tilde{Q}^2}{r^{6}} \right)^{\frac{1}{2}};
\qquad
f_{\rm EH}(r)= 1-\left(\tilde{\Lambda}r^{2} + \frac{2M}{r^{4}}-\frac{\tilde{Q}^2}{r^{8}} \right) 
\end{align}
The horizons of both of these black hole solutions correspond to the solution of the equations $f_{\rm GB}(r)=0$ and $f_{\rm EH}(r)=0$. The existence of three positive real roots of these equations can be easily realized from the Descarte rule of sign applied to the solutions of the above equations. 
\begin{figure}[!htp]
\centering
\subfloat{{\includegraphics[scale=0.22]{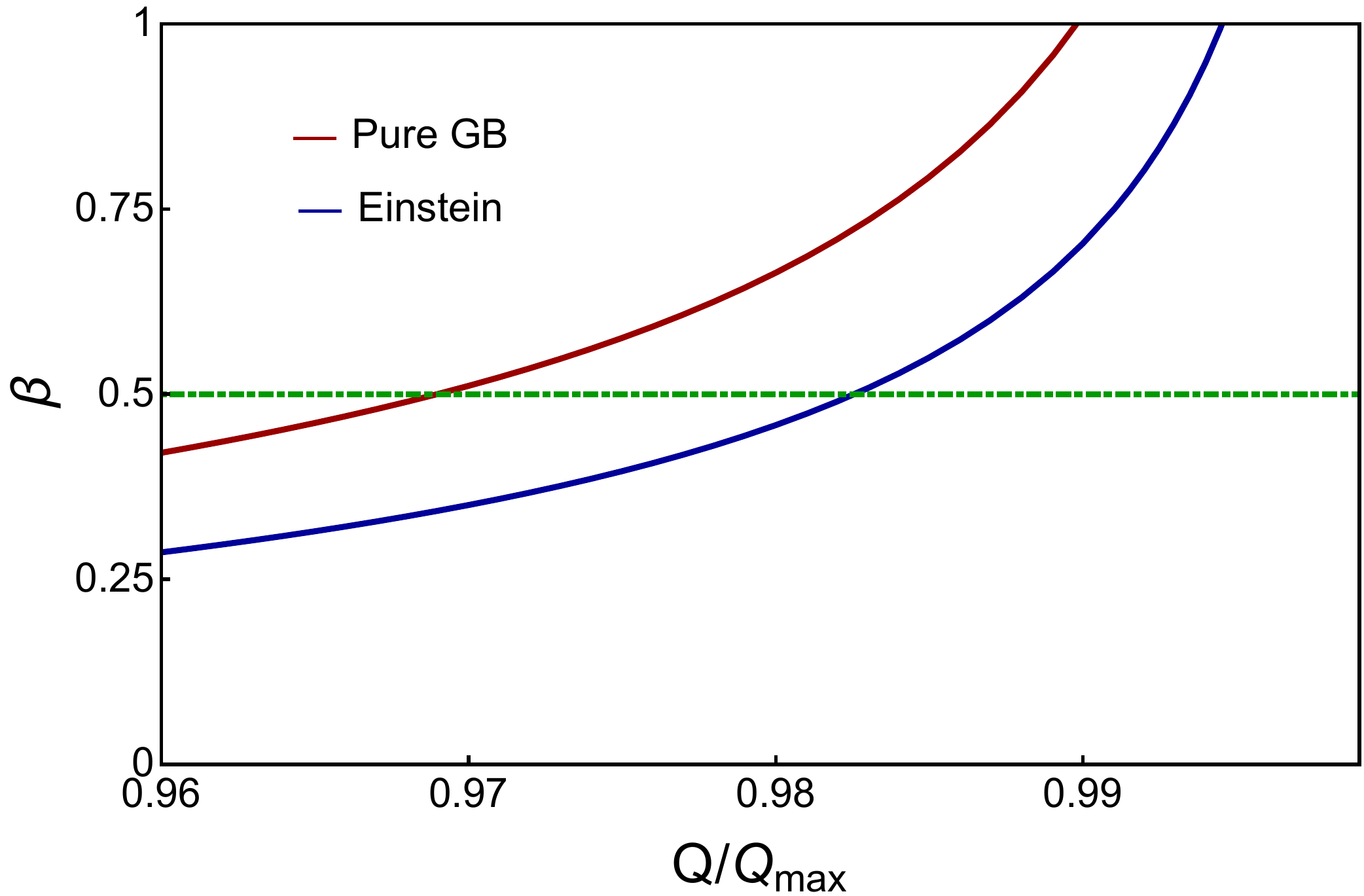} }}~~
\subfloat{{\includegraphics[scale=0.37]{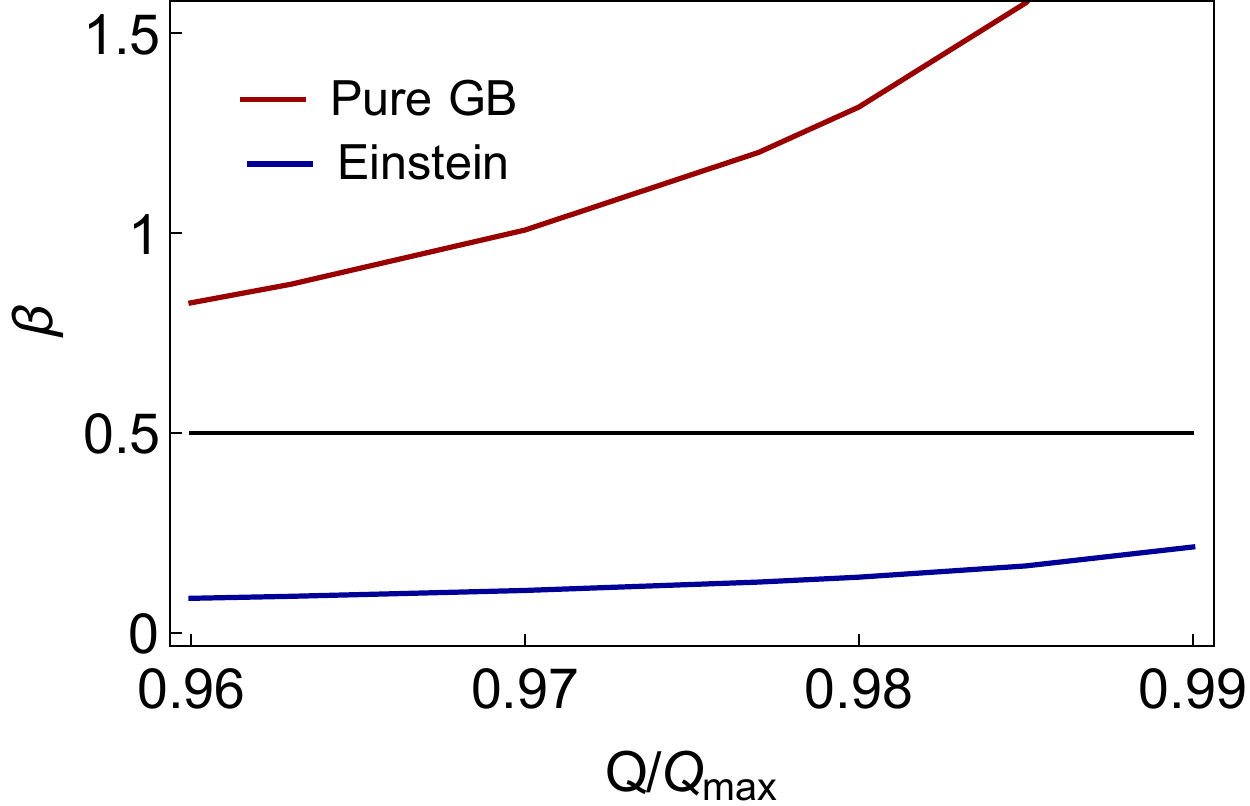} }}~~
\subfloat{{\includegraphics[scale=0.37]{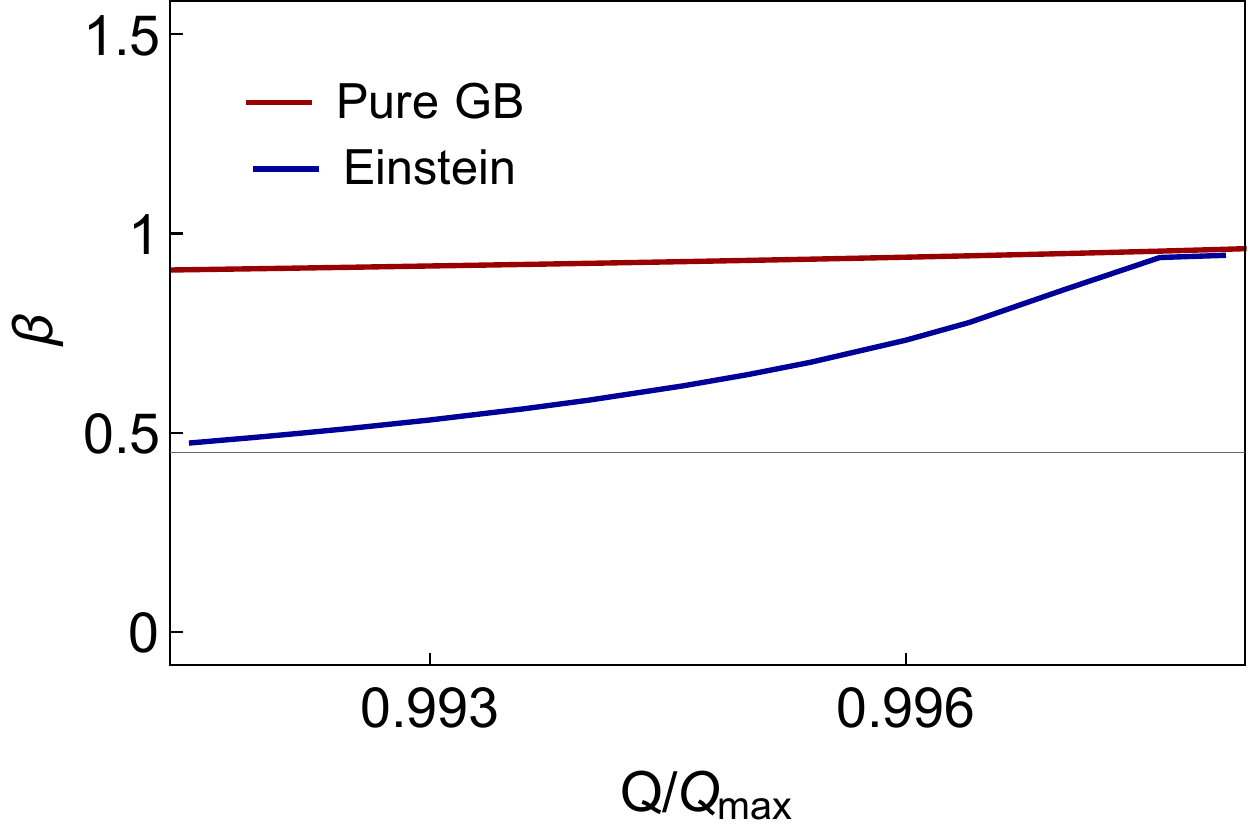} }}
\\
\subfloat{{\includegraphics[scale=0.37]{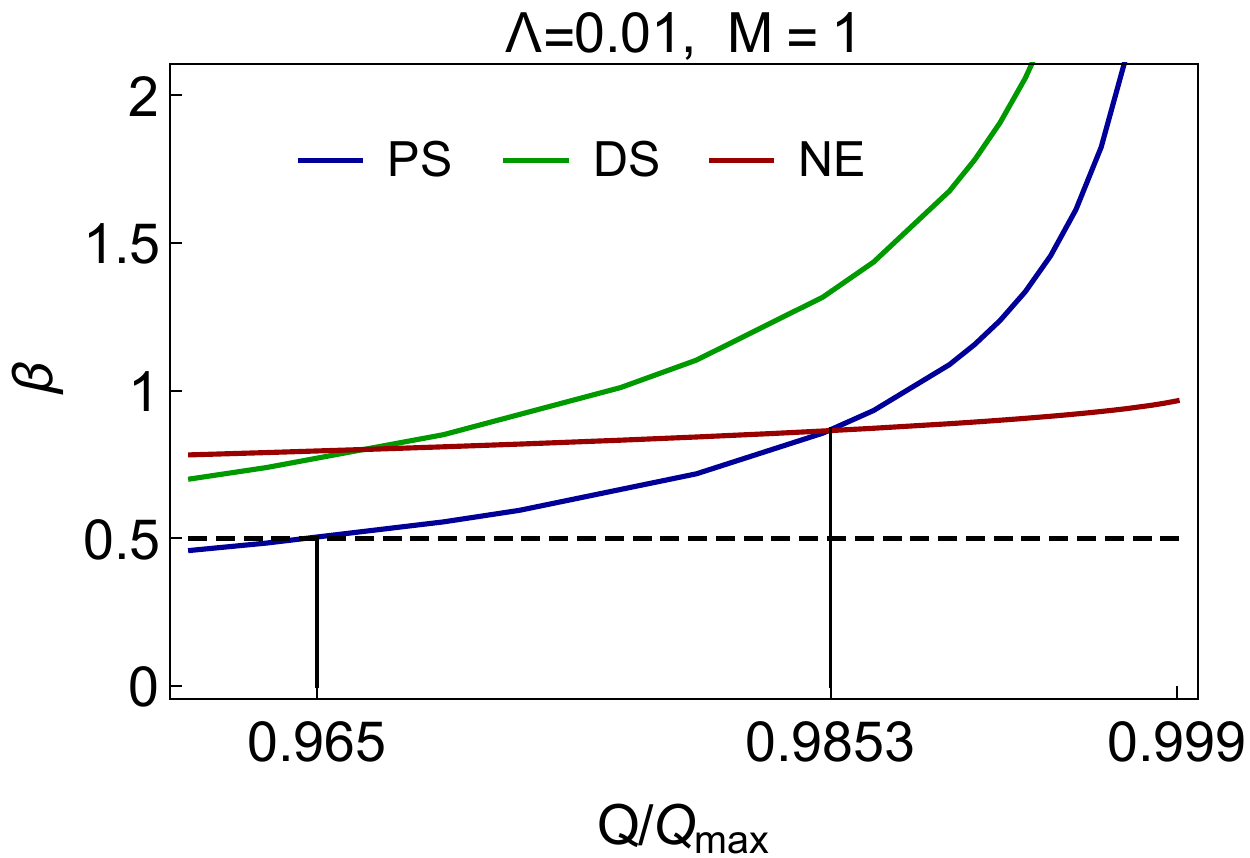} }}~~
\subfloat{{\includegraphics[scale=0.37]{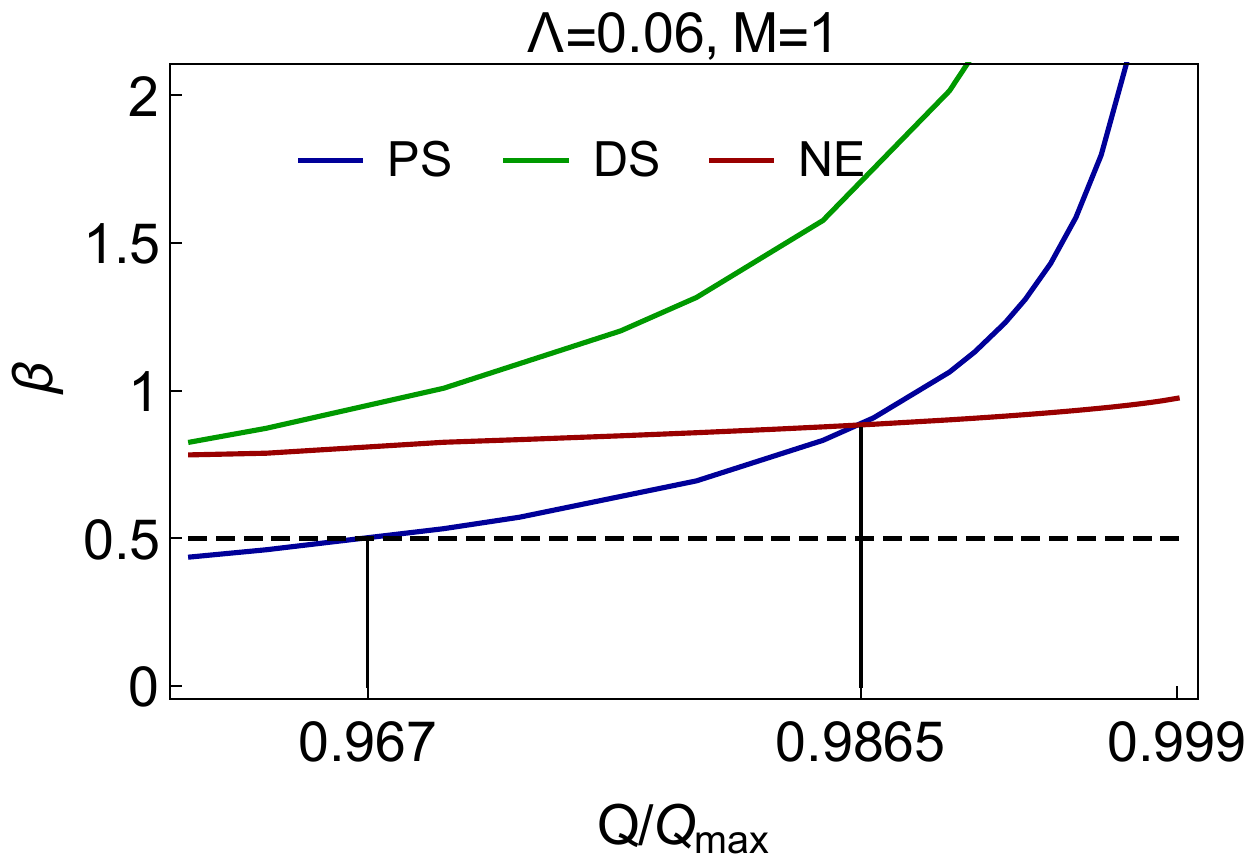} }}~~
\subfloat{{\includegraphics[scale=0.37]{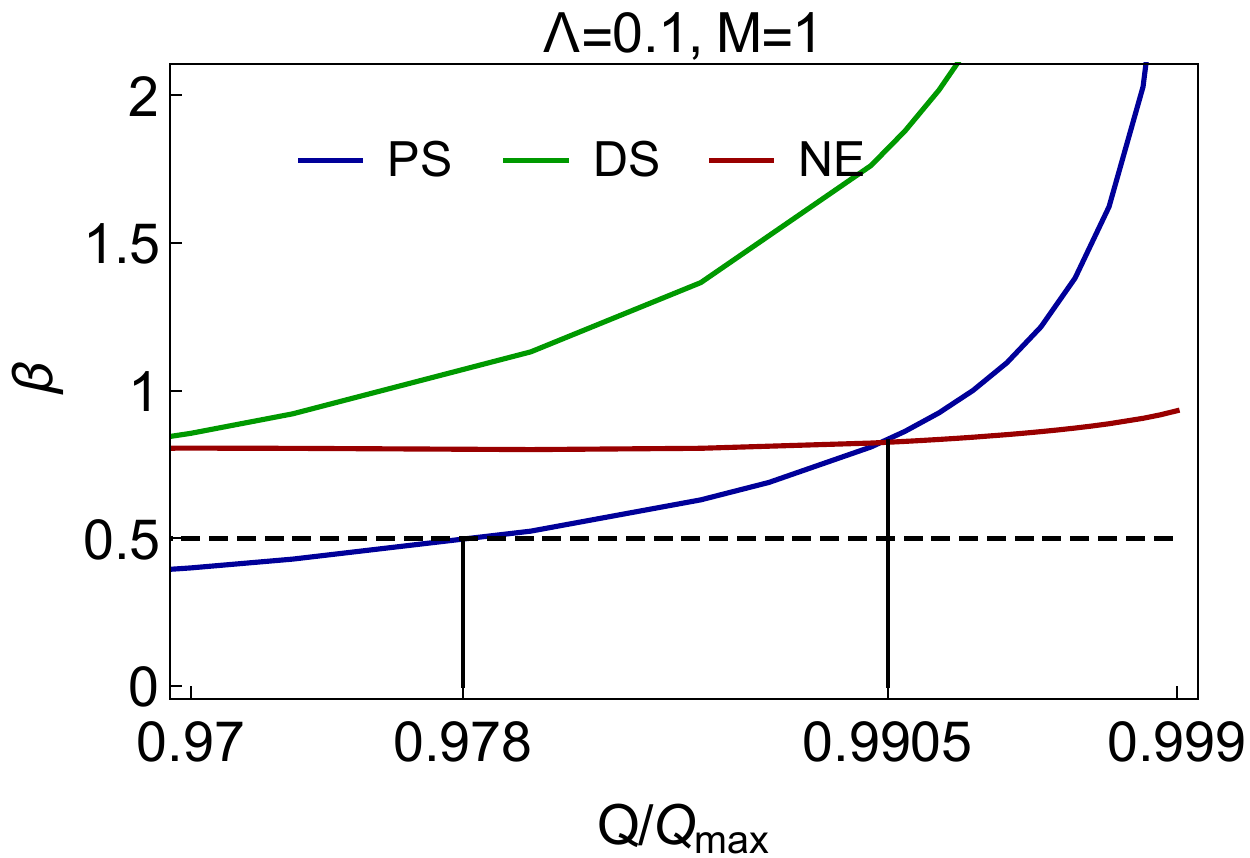} }}

\caption{We have plotted $\{-(\textrm{Im}~\omega_{n})/\kappa_{\rm ch}\}$ (for notational convenience, we have labeled the axis as $\beta$) in the context of pure Gauss-Bonnet gravity as well as Einstein gravity, with the ratio $(Q/Q_{\rm max})$, for the \PS(top left), \DS (top middle) and \NE(top right) modes in the upper row. As evident from the plots of all the three modes, the \scc\ is violated in pure Gauss-Bonnet gravity. Moreover, all these plots demonstrate that the violation of \scc\ is more strong for pure Gauss-Bonnet gravity than the Einstein gravity in $d=7$ dimensions. Here we have taken the cosmological constant to be $\Lambda =0.06$ and the mass being $M=1$. The bottom panel demonstrates the variation of $\{-(\textrm{Im}~\omega_{n})/\kappa_{\rm ch}\}$ for seven dimensional pure Gauss-Bonnet gravity with various choices of the cosmological constant. As the figures clearly demonstrate, the \PS mode always dominate till the near extremal modes take over. Note that the dominant \qnm modes, which is the \NE mode in all the three plots always stays smaller than unity.}
 \label{7d_pure_GB}
\end{figure}

To check the validity of the \scc\ we follow the identical procedure adopted for the case of \EGB gravity in the previous section. We start by computing the quasi-normal mode frequencies and then the relative ratio between the late-time decay rate governed by the imaginary part of the lowest lying \qnm modes and growth at the \ch\ corresponding to a massless scalar perturbation. Following which we have demonstrated the variation of $\{-(\textrm{Im}~\omega_{n})/\kappa_{\rm ch}\}$, whose minima provides an estimation for $\beta$, with the electric charge $(Q/Q_{\rm max})$, where $Q_{\rm max}$ corresponds to the extremal value of the electric charge in \ref{7d_pure_GB}. We have depicted the variation of $\{-(\textrm{Im}~\omega_{n})/\kappa_{\rm ch}\}$ for a given $\Lambda$, for the photon sphere, \DS and \NE modes. As the figures in the upper panel of \ref{7d_pure_GB} explicitly demonstrates, the \scc\ is indeed violated in the pure Lovelock spacetimes as well. Further, as evident from \ref{7d_pure_GB}, the violation of \scc\ in a pure \LL theory occurs at smaller values of $Q/Q_{\rm max}$ in comparison with the similar solution for Einstein gravity in the same spacetime dimensions. Thus we can safely conclude that the violation of \scc\ is more severe in theories involving higher order Lovelock terms than in \gr. For completeness, we have also plotted the variation of $\{-(\textrm{Im}~\omega_{n})/\kappa_{\rm ch}\}$ with $(Q/Q_{\rm max})$ for the three modes, with different choices of the cosmological constant, in the context of pure Gauss-Bonnet black hole in seven spacetime dimensions. As these figures in the bottom panel of \ref{7d_pure_GB} explicitly demonstrates, the \scc\ is indeed violated. In the initial phase, $\beta$ is dominated by the photon sphere modes, while later on the near extremal modes take over. However, the $\beta$ value for the dominating mode never crosses the value unity, which suggests that the field can be extended beyond the Cauchy horizon, with regularity as $H_{\rm loc}^{1}$ function, but not smoother than that. This is consistent with the expectation from general relativistic scenario as well. Thus in this respect the higher curvature gravity follows the footsteps of \gr, even though the parameter space available for violation of \scc\ is larger in higher curvature theories of gravity.  


\section{Conclusion}\label{Section 5}

Given an initial field configuration, predicting what happens to the field in the future through the field equation, is one of the essential features that any well-behaved theory of nature must posses. For the case of \gr, this is ensured by the strong cosmic censorship conjecture, which states that the extension of spacetime metric across the Cauchy horizon keeping the Christoffel symbols as square integrable functions is impossible. However, recently it has appeared that this version of strong cosmic censorship conjecture is seemingly violated for charged black hole spacetimes with a positive cosmological constant. This suggests that the classical fate of an observer is not completely determined from the initial data in \gr. This is a cause for alarm, since it depicts that deterministic nature of \gr\ may break down under certain situations. At the same time one must take cognizance of the fact that \gr\ is only an effective theory and it must be supplemented by higher curvature corrections. Thus there is a tantalizing possibility that \scc\ may be respected when these higher curvature terms are taken into account. 

Following this possibility, we have considered two well known higher curvature theories of gravity and have explored whether \scc\ is respected for charged asymptotically de Sitter black hole solutions in them. In particular, we have studied the fate of \scc\ in the context of a spherically symmetric black hole spacetime in \EGB gravity as well as in pure Gauss-Bonnet theory. We have started by computing the \qnm frequencies in these black hole spacetimes both analytically as well as numerically, which shows good agreement between them (see \ref{table-QNM-GB}). Following which we have determined the minimum of the imaginary part of the \qnm mode frequency as well as surface gravity at the Cauchy horizon, which have helped to study the variation of $\beta$ with the electromagnetic charge $(Q/Q_{\rm max})$, where $Q_{\rm max}$ corresponds to the extremal value of the charge. Surprisingly, we find that \scc\ is violated even in these contexts. Moreover, as the Gauss-Bonnet coupling constant, which characterizes the strength of the higher curvature terms, increases the violation of \scc\ becomes stronger. This conclusion has been achieved by considering all the three families of modes, namely the \PS modes, \DS modes, and \NE modes. Thus our analysis concludes that the addition of higher curvature terms does \emph{not} cure the problem with \scc, rather they lead to an even stronger violation of the \scc. 

To further strengthen our result, we have considered one more type of higher curvature theory, namely, the pure \LL theory of gravity. In particular, we have studied the case of a static, spherically symmetric black hole solution with a positive cosmological constant in seven-dimensional pure Gauss-Bonnet gravity. Following an identical approach to the \EGB case, i.e., by first computing the \qnm modes and then computing $\beta$ for the \PS modes and \NE modes, we see that violation of \scc\ is present even in this context. Furthermore, by comparing this result to that of the pure Einstein gravity in seven dimensions we could conclude that violation of \scc\ is stronger in the case of pure Gauss-Bonnet gravity. Thus our result suggests that the strong cosmic censorship conjecture is violated even when higher curvature terms are included and the violation is stronger than \gr. Moreover, finally the near extremal modes dominate and they reach the value unity asymptotically as $Q$ approaches $Q_{\rm max}$. This suggests that the fields are extendible across the Cauchy horizon as function while belonging to $H_{\rm loc}^{1}$, but not smoother than that. In this respect the results are at per with the case of \gr. As a future outlook, one may consider a similar scenario for rotating black hole solutions in presence of higher curvature and see whether \scc\ is still violated. This will help us to understand the origin of this violation in a better manner. Further, the above result was derived for scalar field and we hope that similar analysis will go through for Dirac and electromagnetic fields, but for gravitational perturbations one needs to be careful due to the presence of higher curvature terms. These we leave for the future. 
\section*{Acknowledgement}

AKM would like to thank IACS, Kolkata and Arizona State University, Tempe, for hospitality where part of the work was carried out. AKM thanks Maulik Parikh, Mostafizur Rahman and Sudipta Sarkar for helpful discussions. AKM is also grateful to IIT Gandhinagar for providing the overseas research fellowship. Research of SC is funded by the INSPIRE Faculty fellowship (Reg. No. DST/INSPIRE/04/2018/000893) from Department of Science and Technology, Government of India.

\bibliography{SCC}

\bibliographystyle{./utphys1}

\end{document}